%% file: adjoint.tex
\documentclass{jfm}

        \input setupAdjoint
        \input defAdjoint

\title[Invariant solutions of the Navier--Stokes equation]
      {An adjoint-based approach for finding invariant
       solutions of Navier--Stokes equations~\footnote{Accepted for publication in the J. Fluid Mech.}}
\author[
M. FARAZMAND
        ]
{
M.\ns F\ls A\ls R\ls A\ls Z\ls M\ls A\ls N\ls D
}
\affiliation{
 Center for Nonlinear Science,
 School of Physics,
 Georgia Institute of Technology,
 Atlanta, GA  30332, USA
}

\begin{document}
\maketitle

We consider the incompressible Navier--Stokes equations with periodic boundary conditions and
time-independent forcing. For this type of flow, we derive adjoint equations whose
trajectories
converge asymptotically to the equilibrium and traveling wave solutions
of the Navier--Stokes equations. Using the adjoint equations,
arbitrary initial conditions evolve
to the vicinity of a (relative) equilibrium at which point a few
Newton-type iterations yield the desired (relative) equilibrium solution. We apply this
adjoint-based method
to a chaotic two-dimensional Kolmogorov flow. A convergence rate of $100\%$ is observed,
leading to the discovery
of $21$ new steady state and traveling wave solutions at Reynolds number $Re=40$.
Some of the new invariant solutions have spatially localized structures
that were previously believed to only exist on domains with large aspect ratios.
We show that one of the newly found
steady state solutions underpins the temporal intermittencies, i.e., high energy
dissipation episodes
of the flow. More precisely, it is shown that each intermittent episode of a generic turbulent
trajectory
corresponds to its close passage to this equilibrium solution.

\section{Introduction}\label{s:intro}
The idea that turbulent fluid flow can be studied in terms of
the invariant solutions of Navier--Stokes equations dates back to
1940s~\citep{hopf48}. Examples of such invariant solutions (also called
`exact coherent states' or `recurrent flows' in fluid dynamics
literature) are
steady states (or equilibria), traveling waves
(or relative equilibria), periodic orbits, relative periodic orbits and
partially hyperbolic tori.
A turbulent flow is then viewed as a walk from the neighborhood of
one invariant solution to the other. If all these solutions are unstable,
the turbulent trajectory never settles down and its itinerary becomes
desperately complex.

Initially a mathematical endeavor, this view has been put to practice over the last
15 years in many experimental and numerical studies, providing insights
that are beyond the reach of purely statistical descriptions of turbulence
(see \rf{KUV12} and \rf{C13}, for reviews). 

The invariant solutions of Navier--Stokes equations often exhibit
complex spatiotemporal behaviors, and hence analytic expressions for them
are unavailable. Their numerical computation was first
explored in the pioneering works of\rf{marcus87} and\rf{N90}.
Nagata found the first non-trivial equilibrium and traveling wave solutions
of the plane Couette flow~\citep{N90,N97}.
Nagata's approach to finding these solutions is the following.
Through a finite truncation, the invariant solutions are formulated as the zeros of
a large system of nonlinear algebraic equations $\vc F(\vc a)=\vc 0$. The solution $\vc a$ is
then found through Newton--Raphson iterations. Depending on the type of the
discretization, the finite vector $\vc a$ contains either
the Galerkin coefficients or the collocation values.

Nagata's approach underlies the state-of-the-art methods for finding invariant solutions
of fluid flow
(see, e.g.,\rf{W98,KawKida01,DV04,HGC08,ACHKW11}). Such Newton-type approaches suffer from two
major drawbacks:
\begin{enumerate}\renewcommand{\theenumi}{\roman{enumi}}
\item Obtaining the Newton direction $\delta\vc a$ requires solving the linear system of equations
$\bnabla\vc F(\vc a)\delta\vc a=-\vc F(\vc a)$.
Even for moderate Reynolds numbers, this linear equation
is too large to be solved accurately (within numerical precision) in reasonable computation time.
\item The convergence of the Newton iterations is not generally guaranteed. The iterations converge
only if very good initial guesses are provided.
\end{enumerate}

As regards item (i), the linear system is typically too large for the
matrix $\bnabla\vc F(\vc a)$ to be even
explicitly formed (due to memory considerations). To address this issue,
 the generalized minimal residual (GMRES) method is used to
approximate the solution of the linear system~\citep{gmres}. The GMRES method replaces the
exact Newton direction with its least square approximation within a Krylov subspace.
Forming this subspace only requires the action of the matrix $\bnabla\vc F(\vc a)$ on certain
vectors, hence avoiding the formation of the matrix itself.
The resulting method is often referred to as the Newton--GMRES iterations.
Although computationally feasible, the Newton--GMRES iterations are still
a formidable numerical undertaking in terms of implementation and computation time.

As mentioned in item (ii) above, even when the exact Newton descent direction $\delta \vc a$
is known, the iterations are not guaranteed to converge. Newton's domain of convergence
in the context of Navier--Stokes equations is typically small,
demanding a very good initial guess for the iterations to converge~\citep{W97}. The Newton
iterations, therefore, are very effective
for bifurcation analysis where the invariant solutions bifurcate from known
solutions~\citep{Tuck00,W03,K12}. In this case, the known solution (corresponding to the
old bifurcation parameter value) is used as the initial guess for the Newton iterations.
Even then, the method might fail to determine solutions far from the bifurcation point.

In practice, good initial guesses for Newton iterations are obtained through heuristic methods,
such as close recurrences of a turbulent trajectory~\citep{pchaot,kawahara06,DV04,CviGib10}.
Such heuristics lack a
solid mathematical basis. As a result, many important invariant solutions may and will remain undiscovered
(see Section~\ref{sec:interm}, below).

\cite{DV04} used the \emph{hook step}
together with the Newton--GMRES
method which significantly improved the convergence of the iterations. The
underlying idea of the hook step is to
approximate the Newton direction within the Krylov subspace with the constraint that
$\delta \vc a$ belongs to a `trust region' such that
$|\delta \vc a|$ is smaller than some prescribed value $\epsilon$. The
constraint $|\delta\vc a|<\epsilon$ ensures that the linearization
$\bnabla\vc F(\vc a)\delta\vc a$ involved in Newton iterations is a valid approximation~\citep{DS}.
The resulting method, i.e. the Newton--GMRES-hook iterations, is state-of-the-art
in computing invariant solutions in the context of fluid flow.
While improving the convergence chance of
the Newton--GMRES iterations, the hook step still requires a good
initial guess.

Such drawbacks have slowed down research progress in transitional and moderate Reynolds
number turbulence.
As the study of fluid turbulence through its invariant solutions reaches a mature state, it is time
to revisit the methods by which these invariant solutions are found. 
Better methods should ideally scale linearly (in the sense of computational complexity) 
with the number of degrees of freedom and be
universally convergent. The term \emph{universally convergent} implies that every initial guess converges
to some solution. Universally convergent methods, however, do not immediately guarantee the
identification of all possible solutions from a finite set of initial guesses.

The main goal of the present paper is to propose one such new direction.
We develop an adjoint-based method for computing steady state and
traveling wave solutions of incompressible Navier--Stokes equations with periodic
boundary conditions. In particular, adjoint partial differential equations (PDEs) are derived whose 
trajectories converge to
(relative) equilibria of the Navier--Stokes equations.

Adjoint equations appear naturally in optimal control, where the governing equations
act as constraints (see\rf{gunzburger}, for a survey of applications in fluid dynamics). 
\cite{PK10} and\rf{henningson2011} used adjoint-based optimization
to find the optimal route to turbulence in transitional shear flows (also see\rf{KPW14} for a review).
To investigate the regularity of the Naveir--Stokes equations,\rf{AP14} use
adjoint-based optimization to find the initial conditions that result in maximal palinstrophy growth.

To the best of our knowledge, adjoint-based methods have not been used to 
find invariant solutions of the Navier--Stokes equations. They have, however, been 
utilized in the context of nonlinear wave equations.
For instance,\rf{Ambrose10} formulate the
time-periodic solutions of a unidirectional water wave equation
as the minima of an appropriate functional. The minima are found 
iteratively by
a `steepest descent' method.
At each iteration,
the descent direction is obtained as the solution of a backward-time adjoint PDE (also see\rf{Ambrose10-2}).
\cite{yang07} realize that, for equilibrium solutions, the descent direction can be expressed explicitly.
Namely, they show that the solitary solutions of nonlinear wave equations coincide with the asymptotically stable equilibria of 
an adjoint PDE.

The case of
Navier--Stokes equations is more involved due to
the presence of nonlocal pressure
gradients enforcing incompressibility. This calls for a special treatment
as presented here. We also extend the adjoint-based approach to finding
traveling wave solutions with a priori unknown wave speeds.

Recently,\rf{CK13} and\rf{LK14} carried out the most exhaustive search for invariant solutions
of a chaotic two-dimensional Kolmogorov flow using Newton-GMRES-hook iterations.
Here, we apply our adjoint-based method to the same Kolmogorov flow and discover several
new steady state and traveling wave solutions. We show that some of these new
solutions underpin the temporal intermittencies associated with
short-lived, rapid growth of energy dissipation. 

This paper is organized as follows. The preliminary material is reviewed in
Section~\ref{sec:prelim}, in a rather general setting. The explicit form of the
adjoint equations for Navier--Stokes equations are presented in
Section~\ref{sec:adj_NS}. We test the adjoint-based approach in Section~\ref{sec:test}
on a two-dimensional Kolmogorov flow
and also carry out a thorough comparison with Newton--GMRES-hook iterations.
In Section~\ref{sec:interm}, the temporal intermittency of the Kolmogorov flow is
studied in terms of its invariant solutions. Finally, Section~\ref{sec:conclude} contains
our concluding remarks and an outline of future research directions.

\section{Preliminaries: Newton descent vs. adjoint descent}\label{sec:prelim}
In this section, we review Newton and adjoint descent methods
in a general framework.
We find this exposition helpful for
understanding the Navier--Stokes adjoint
equations presented in Section~\ref{sec:adj_NS}.

Consider the partial differential equation (PDE)
\beq
\partial_t \vc u = \vc F(\vc u),
\label{eq:pde}
\eeq
where the real valued vector field
$\vc u(\vc x,t)$ is a function of space $\vc x$ and time $t$, and
$\vc F$ is a nonlinear differential operator acting on an inner product function space
$\mathcal H$. We seek equilibrium (or steady state) solutions of this PDE, i.e.,
time-independent vector fields $\vc u=\vc u(\vc x)$ such that $\vc F(\vc u)=\vc 0$.

The finite dimensional counterpart of~\eqref{eq:pde} is the system of ordinary differential
equations (ODEs) $\mbox d\vc a/\mbox d t=\vc F(\vc a)$ where $\vc a\in\mathbb R^d$ and
$\vc F:\mathbb R^d\to\mathbb R^d$. While our derivations are in the infinite-dimensional setting,
we will occasionally invoke this finite-dimensional analogue to elaborate the ideas.

We define the `weighted' inner product
\beq
\langle \vc u,\vc u'\rangle_{\mathcal A}=\langle \vc u,\mathcal A\,\vc u'\rangle_{L^2},
\label{eq:innp_H-1}
\eeq
for any two functions $\vc u,\vc u'\in\mathcal H$,
where $\mathcal A$ denotes a real-valued, positive-definite, self-adjoint (with respect to the $L^2$
inner product $\langle\cdot,\cdot\rangle_{L^2}$) operator.
The prime signs should not be confused with derivatives; we use them here
to distinguish different functions.
The inner product~\eqref{eq:innp_H-1} defines a natural norm given by
\beq
\|\vc u\|_{\mathcal A}=\sqrt{\langle \vc u,\vc u\rangle_{\mathcal A}},
\label{eq:normA}
\eeq
for $\vc u\in\mathcal H$.
The choice of $\mathcal A$ is somewhat arbitrary and 
should be judiciously chosen for a specific
problem. For the Navier--Stokes equation, for instance, we will use an operator
which is closely related to the inverse of the Laplacian (see Section~\ref{sec:adj_NS}).
For this exposition, however, we set $\mathcal A$ to identity, which renders~\eqref{eq:innp_H-1}
as the usual $L^2$ inner product.

With this setting, we seek a second PDE
\beq
\partial_\tau\vc u = \vc G(\vc u),
\label{eq:descend_inf}
\eeq
such that its solutions
converge to the equilibrium solutions of~\eqref{eq:pde} as the fictitious
time $\tau$ tends to infinity.  To ensure this convergence, the differential operator $\vc G$ is
designed in a specific way such that
$$\|\vc F(\vc u(\tau))\|_{L^2}\to 0,\quad \mbox{as}\quad\tau\to\infty,$$
where $\vc u(\tau)= \vc u(\cdot, \tau)$ denotes a solution of \eqref{eq:descend_inf}.
Newton and adjoint descents correspond to two different choices of the operator
$\vc G$.

The finite-dimensional analogue of~\eqref{eq:descend_inf} is the fictitious-time ODE
$\mbox d\vc a/\mbox d\tau=\vc G(\vc a)$
with $\vc G:\mathbb R^d\to\mathbb R^d$.

\subsection{Newton descent}\label{sec:descend_inf}
The main observation here is that discrete Newton iterations are
an explicit Euler approximation of a continuous fictitious-time dynamical
system~\citep{saupe88}.
This well-known fact is
rarely mentioned in the literature, prompting
\cite{smale81} to write ``\textit{This geometry is based on an idea which is known,
but not usually explicated. Namely, Newton's method for
solving $f(z)=0$ is an Euler approximation to the ordinary differential equation...}".
Here, we first derive the continuous fictitious-time Newton method in the infinite-dimensional setting
and then show how the discrete Newton iterations follows from there.

The norm $\|\vc F
(\vc u)\|_{L^2}$ evolves along the trajectories of \eqref{eq:descend_inf} according to
\beq
\partial_\tau \|\vc F
(\vc u)\|^2_{L^2}=2\langle \pmb{\mathcal L}(\vc u;\vc G(\vc u)),\vc F(\vc u)\rangle_{L^2},
\label{eq:norm_ev}
\eeq
where the linear
operator $\pmb{\mathcal L}(\vc u;\cdot)$ is the G\^ateaux derivative of $\vc F$ evaluated
at the state $\vc u$, and is defined by
\beq
\pmb{\mathcal L}(\vc u;\vc u'):=\lim_{\epsilon\to 0}\frac{\vc F(\vc u+\epsilon\vc u')-\vc
	F(\vc u)}{\epsilon},\quad \forall\,\vc u' \in\mathcal H.
\eeq

Requiring  $\vc G$ to satisfy
\beq
\pmb{\mathcal L}(\vc u;\vc G(\vc u))=-\vc F(\vc u),
\label{eq:G_newton_inf}
\eeq
one obtains the continuous fictitious-time Newton descent in infinite dimensions.
Substituting this expression in~\eqref{eq:norm_ev}, we obtain
\beq
\partial_\tau \|\vc F
(\vc u(\tau))\|^2_{L^2}=-2 \|\vc F
(\vc u(\tau))\|^2_{L^2},
\eeq
which has the exact solution $\|\vc F(\vc u(\tau))\|_{L^2}=\|\vc F(\vc u(0))\|_{L^2}\exp(-\tau)$.
This shows that, as long as~\eqref{eq:G_newton_inf} has
a solution, $\|\vc F(\vc u(\tau))\|_{L^2}$ decays to zero exponentially fast
along the solutions of~\eqref{eq:descend_inf}. In other words, the continuous fictitious-time
Newton method converges to an equilibrium solution of~\eqref{eq:pde} for
almost all initial conditions~\citep{saupe88,CvitLanCrete02,lanVar1}.

For a given state $\vc u$, Eq.~\eqref{eq:G_newton_inf} is a PDE that can in principle be solved
for $\vc G$. In practice, it is approximated through some finite truncation
to take the form $\bnabla\vc F(\vc a)\vc G(\vc a)=-\vc F(\vc a)$, with $\vc a$ being the
Galerkin coefficients.
The solution $\vc G$ of this large, but finite-dimensional, linear system is often
approximated by some variant of
generalized minimal residual (GMRES) iterations~\citep{Trefethen97}.

Furthermore, the PDE \eqref{eq:descend_inf} is discretized in time to yield the explicit Euler step
\begin{equation}
\vc u_{i+1}=\vc u_i +\delta\tau\, \vc G(\vc u_i),
\label{eq:Eul_it}
\end{equation}
with $0<\delta\tau\leq 1$. The standard Newton iterations correspond to $\delta\tau=1$, while
\emph{damped} Newton iterations adjust $\delta\tau$ in order to ensure that the $L^2$ norm
decreases~\citep{boyd04}.

The above discrete iterations do not guarantee the global
convergence that the continuous fictitious-time Newton descent does~\citep{saupe88}. As mentioned
in the Introduction,
the convergence of Newton iterations is only guaranteed if the initial guess
is sufficiently close to an equilibrium. In fact, if such a close initial guess is
available, the convergence to the equilibrium is super-exponential~\citep{Deuf11}.

Anecdotal evidence suggests that, in the context of fluid flow,
the basin of attraction of Newton iterations is rather small~\citep{W97,DV04}.
This domain can
be enlarged by choosing a higher order scheme for temporal discretization of~\eqref{eq:descend_inf}.
This is, however, computationally too demanding
for large dimensional systems, such as turbulent fluid flow.

\subsection{Adjoint descent}
The adjoint descent corresponds to an alternative choice of $\vc G$ that  may be expressed
analytically, thus avoiding the
approximation involved in the Newton-GMRES iterations. Moreover, the adjoint direction can be
evaluated at
a relatively low computational cost, rendering a higher order time discretization of
\eqref{eq:descend_inf} feasible.

To express the adjoint direction, note that \eqref{eq:norm_ev} can be written as
\beq
\partial_\tau \|\vc F
(\vc u)\|^2_{L^2}=2\langle\vc G(\vc u), \pmb{\mathcal L}^\dagger(\vc u;\vc F(\vc u))\rangle_{L^2},
\label{eq:norm_ev2}
\eeq
where the adjoint $\pmb{\mathcal L}^\dagger(\vc u;\cdot)$ is the linear operator satisfying
\beq
\langle \pmb{\mathcal L}(\vc u;\vc u'),\vc u''\rangle_{L^2} = \langle \vc
u',\pmb{\mathcal L}^\dagger(\vc
u;\vc u'')\rangle_{L^2},\quad \forall\, \vc u,\vc u',\vc u''\in L^2.
\label{eq:adj_def}
\eeq

Requiring
\beq
\vc G(\vc u)=-\pmb{\mathcal L}^\dagger(\vc u;\vc F(\vc u)),
\label{eq:adjoint_dir_inf}
\eeq
ensures the descent of the norm $\|\vc F(\vc u(\tau))\|_{L^2}$ along the trajectories of
\eqref{eq:descend_inf}, since
\beq
\partial_\tau \|\vc F
(\vc u)\|^2_{L^2}=-2\|\pmb{\mathcal L}^\dagger(\vc u;\vc F(\vc u))\|_{L^2}^2\leq 0.
\label{eq:adj_rate_descent}
\eeq
We refer to $\vc G$ given by~\eqref{eq:adjoint_dir_inf} as the \emph{adjoint direction}.
The finite-dimensional analogue of this adjoint direction is
$\vc G(\vc a)=-\left[\bnabla\vc F(\vc a)\right]^\top\vc F(\vc a)$, where $\top$ denotes
the usual matrix transposition.

The downside of using the adjoint descent is that, unlike the continuous fictitious-time
Newton descent, equation~\eqref{eq:adj_rate_descent} does not guarantee an
exponential decay of the $L^2$ norm. In fact, we observe
a relatively slow decay in Section~\ref{sec:hybrid}, and propose a
hybrid adjoint-Newton algorithm to rectify this drawback.

In Section~\ref{sec:adj_NS}, we derive the explicit form of the adjoint operator for the
incompressible Navier--Stokes equations.

\section{Adjoint descent for the Navier--Stokes equations}\label{sec:adj_NS}
Consider the incompressible Navier--Stokes equation in non-dimensional variables,
\begin{subequations}
\beq
\partial_t \vc u = -\vc u\cdot\bnabla\vc u -\bnabla p +\nu\Delta\vc u+\vc
f,
\label{eq:NS1}
\eeq
\beq
\bnabla\cdot\vc u=0,
\eeq
\label{eq:NS}%
\end{subequations}
defined on the spatial domain $\vc x=(x_1,x_2,x_3)\in\mathcal D= [0,L_1]\times [0,L_2]\times
[0,L_3]$ with
periodic
boundary conditions. The parameter $\nu$ is the inverse of the Reynolds number, $\nu=Re^{-1}$.
We assume that the
time-independent forcing term $\vc f=\vc f(\vc x)$ is divergence-free,
$\bnabla\cdot\vc f=0$.

The goal is to find an adjoint PDE such that, along its solutions $(\vc u(\tau),p(\tau))$, the
norm of the right-hand-side of~\eqref{eq:NS1} decays to zero monotonically, while ensuring that
$\bnabla\cdot\vc u=0$ for all $\tau$. We derive the adjoint equations for the general norms
of the type~\eqref{eq:normA}. In Section~\eqref{sec:H-1}, we make a specific choice for the norm
that is used in the subsequent computations.

\subsection{The adjoint descent equation for equilibria}
Define
\beq
\vc F_{\vc 0}(\vc u)=
-\vc u\cdot\bnabla\vc u -\bnabla p +\nu\Delta\vc u+\vc f,
\eeq
which is the right-hand-side of Eq.~\eqref{eq:NS1}.
The reason for using the subscript $\vc 0$ in $\vc F_{\vc 0}$ becomes
clear in
Section~\ref{sec:tw}, where we derive the adjoint equation for traveling waves.
We seek steady states $\vc u$ such that $\vc F_{\vc 0}(\vc u)=\vc 0$.

To this end, we derive an adjoint PDE, such that, along
its trajectories $\vc u(\tau)$, the norm $\|\vc F_{\vc 0}(\vc u(\tau))\|_{\mathcal A}$
decreases monotonically and converges to zero asymptotically.
Our choice of the norm does not compromise the accuracy of the
resulting equilibrium solutions of the Navier--Stokes equations: due to the positive-definiteness
of the operator $\mathcal A$, $\|\vc F_{\vc 0}(\vc u)\|_{\mathcal A}= 0$ if and only if
$\vc F_{\vc 0}(\vc u)=\vc 0$

Leaving the derivation to Appendix~\ref{app:proof_adj}, the adjoint descent equations read
\begin{subequations}
\beq
\partial_\tau\vc u = -\left\{\left[\bnabla\tilde{\vc u}''+(\bnabla\tilde{\vc u}'')^\top\right]\vc
u-\bnabla p''+\nu\Delta \tilde{\vc u}''\right\},
\label{eq:adjPDE_NS_1} %
\eeq
\beq
\bnabla\cdot\vc u''=0,\quad \bnabla\cdot\vc u=0,
\label{eq:adjPDE_NS_2} %
\eeq
\label{eq:adjPDE_NS}%
\end{subequations}
where $\vc u''=\vc F_{\vc 0}(\vc u)$, i.e.,
\beq
\vc u'' = -\vc u\cdot\bnabla\vc u -\bnabla p +\nu\Delta\vc u+\vc
f,
\label{eq:u''}
\eeq
and we use the shorthand notation $\tilde{\vc u}'':=\mathcal A\vc u''$.

Two pressure-type terms $p$ and $p''$ appear in the adjoint equation. As in the case
of the Navier--Stokes equation, the pressure terms play the role of the Lagrange
multipliers,
enforcing the divergence-free constraints on $\vc u$ and $\vc u''$
(cf. Appendix~\ref{app:proof_adj}).
Taking the divergence of equation~\eqref{eq:u''} we get
\beq
\Delta p = -\bnabla\cdot[\vc u\cdot\bnabla\vc u],
\label{eq:p}
\eeq
enforcing $\bnabla\cdot\vc u''=0$. Similarly, taking the divergence
of~\eqref{eq:adjPDE_NS_1} yields
\beq
\Delta p'' = \bnabla\cdot\left\{\left[\bnabla\tilde{\vc u}''+(\bnabla\tilde{\vc u}'')^\top\right]\vc
u\right\},
\label{eq:p''}
\eeq
ensuring  $\bnabla\cdot\vc u=0$.

Due to the periodic boundary conditions, these
divergence-free constraints can be easily enforced by a projection operator in the
spectral space.
In Appendix~\ref{app:adj_fs}, we derive the Fourier space
representation of the adjoint equations~\eqref{eq:adjPDE_NS}. This spectral representation
closely resembles that of the Navier--Stokes equations. Therefore, an existing pseudo-spectral code
for the Navier--Stokes equations can also be used, with straightforward modifications,
to solve the adjoint equations. Furthermore, the adjoint equations~\eqref{eq:adjPDE_NS}
enjoy the same degree of parallelizability as the Navier--Stokes equations.

\subsection{Adjoint descent for the traveling waves}\label{sec:tw}
In the absence of forcing, the Navier--Stokes equation~\eqref{eq:NS}
is invariant under Galilean translations. This symmetry allows
for the existence of traveling wave (or relative equilibrium) solutions. A traveling wave has the
general form $\vc u(\vc x,t)=\vc u(\vc x-\vc c t)$, where $\vc c=(c_1,c_2,c_3)\in\mathbb R^3$
is a constant wave velocity. The forcing term $\vc f$ may break the translational
symmetry in one or more directions, in which case, the wave
speed $c_i$ corresponding to the symmetry-broken
direction $x_i$ is identically zero.

Substituting the traveling wave solution $\vc u(\vc x-\vc c t)$ into the
Navier--Stokes equation~\eqref{eq:NS}, gives the time-independent equations
\beq
-\vc u\cdot\bnabla\vc u -\bnabla p +\nu\Delta\vc u+\vc f+
\vc c\cdot\bnabla \vc u=\vc 0, \quad
\bnabla\cdot \vc u=0.
\eeq

Therefore, finding traveling wave solutions to the Navier--Stokes equation amounts to finding
the kernel of the operator
\beq
\vc F_{\vc c}(\vc u)=
-\vc u\cdot\bnabla\vc u -\bnabla p +\nu\Delta\vc u+\vc f+\vc c\cdot\bnabla \vc u.
\eeq

If the wave velocity $\vc c$ is known, the solutions to
$\vc F_{\vc c}(\vc u)=\vc 0$ can be found through
an adjoint descent equation similar to~\eqref{eq:adjPDE_NS}. In general, however, the wave
velocity $\vc c$ is unknown. To address this general case, we allow $\vc c$ to be
a function of the fictitious time $\tau$, and enforce its derivative
$\dot{\vc c}=\mathrm d\vc c/\mathrm d\tau$ to change in such a way that
$\|\vc F_{\vc c(\tau)}(\vc u(\tau))\|_{\mathcal A}$ decreases monotonically to zero
as $\tau$ increases.

Leaving the details to Appendix~\ref{app:adj_tw}, we obtain the set of adjoint descent equations
\begin{subequations}
\beq
\partial_\tau\vc u = -\left\{\left[\bnabla\tilde{\vc u}''+(\bnabla\tilde{\vc u}'')^\top\right]\vc
u-\bnabla p''+\nu\Delta \tilde{\vc u}''\right\}+\vc c\cdot\bnabla\tilde{\vc u}'',
\label{eq:adjPDE_tw-1}
\eeq
\beq
\frac{\mathrm d\vc c}{\mathrm d \tau}=-\int_{\mathcal D}\left(\bnabla\vc u\right)^\top\tilde{\vc
u}''\;\mathrm{d}\vc x,
\label{eq:dot_c}
\eeq
\beq
\bnabla\cdot\vc u''=0,\quad \bnabla\cdot\vc u=0,
\eeq
\label{eq:adjPDE_NS_H-1_tw}%
\end{subequations}
where $\vc u''=\vc F_{\vc c}(\vc u)$, i.e.,
\beq
\vc u''=-\vc u\cdot\bnabla\vc u -\bnabla p +\nu\Delta\vc u+\vc f+\vc c\cdot\bnabla \vc u,
\label{eq:u''_tw}
\eeq
and $\tilde{\vc u}''=\mathcal A\vc u''$. The pressure-type functions $p$ and $p''$
satisfy equations~\eqref{eq:p} and~\eqref{eq:p''},
respectively, with $\vc u''$ given by~\eqref{eq:u''_tw}.
The spectral representation of the adjoint equations~\eqref{eq:adjPDE_NS_H-1_tw} is
given in Appendix~\ref{app:adj_fs}.

The adjoint descent for traveling waves includes the evolution equation~\eqref{eq:dot_c}
for the unknown wave velocity $\vc c$.
If the symmetry in $x_i$ direction is broken due to the forcing term $\vc f$, the corresponding
wave speed $c_i$ is identically zero and the equation for $\dot c_i$ is eliminated
from~\eqref{eq:dot_c}.

\subsection{Choice of the norm}\label{sec:H-1}
The conventional choice for the norm $\|\cdot\|_{\mathcal A}$ is the $L^2$ norm, corresponding
to $\mathcal A$ being the identity operation. We find,
however, that the resulting equations are stiff, requiring very small
time-steps for their numerical integration. This is in line with a similar observation
made by\rf{yang07} in the context of nonlinear wave equations.
Here we choose the $H^{-1}$ norm that renders the
adjoint PDEs less stiff by damping the higher Fourier modes.

This choice is somewhat
arbitrary, but we find it to work very well for the Kolmogorov flow presented in
Section~\ref{sec:kolm}. Moreover, the implementation of the resulting equations
in the Fourier space is straightforward.

The $H^{-1}$ can be computed as follows. Let $\vc q=(\vc u,p)$ be the
velocity-pressure pair and define
the operator $\mathcal A$ through its action in
the Fourier space,
\beq
\widehat{\mathcal A\vc q}(\vc k) = \frac{\widehat{\vc q}(\vc k)}{1+|\vc k|^2},
\label{eq:Afs}
\eeq
where $\widehat{\vc q}$ denotes the Fourier transform of $\vc q$ and $\vc
k=(2\pi k_1/L_1,2\pi k_2/L_2,2\pi k_3/L_3)$, with $k_i\in\mathbb Z$ being the
wavenumber.
Note that the operator $\mathcal A$ commutes with the
divergence operation, i.e., $\mathcal A (\bnabla\cdot \vc u)=\bnabla\cdot\mathcal A\vc u=0$.

With this choice and using Parseval's identity,
the inner product~\eqref{eq:innp_H-1} can be written explicitly in the Fourier space as
\beq
\langle \vc q,\vc q'\rangle_{H^{-1}}=\sum_{\vc k}\frac{\widehat{\vc q}(\vc k)\cdot
\widehat{\vc q}'(\vc k)^\ast}{1+|\vc k|^2},
\label{bogusTime}
\eeq
where $\ast$ denotes complex conjugation. This inner product induces the $H^{-1}$ norm
$\|\vc q\|_{H^{-1}}^2=\langle \vc q,\vc q\rangle_{H^{-1}}$.

The operator $\mathcal A$ appears in the adjoint equations
\eqref{eq:adjPDE_NS} and~\eqref{eq:adjPDE_NS_H-1_tw} through
$\tilde{\mathbf u}''=\mathcal A\mathbf u''$,
which can be readily computed in the Fourier space through
Eq.~\eqref{eq:Afs}.

\section{Equilibria and traveling waves for Kolmogorov flow}\label{sec:test}
We test the performance of the adjoint descent equations on a
two-dimensional Kolmogorov flow.
This flow (with identical geometry and parameters as the ones
chosen here) was studied thoroughly by\rf{CK13},
providing a basis for comparison with our results.

In Section~\ref{sec:hybrid}, we introduce the hybrid adjoint-Newton method which requires
the Newton--GMRES-hook (or NGh, for short) iterations. The NGh iterations
are explained in detail by\rf{CK13}.
Our implementation and choice of parameters are
identical to theirs with the exception that we use the velocity formulations of
the Kolmogorov flow, while\rf{CK13} use the vorticity formulation.
We chose to use the velocity formulation since the adjoint equations~\eqref{eq:adjPDE_NS}
and~\eqref{eq:adjPDE_NS_H-1_tw} are derived for the general three-dimensional flow, and hence
are in the velocity-pressure form.

\subsection{Kolmogorov flow}
\label{sec:kolm}
Considered the forced Navier--Stokes equation \eqref{eq:NS} on the two-dimensional torus
$\vc x=(x_1,x_2)\in\mathbb T^2=[0,2\pi]\times[0,2\pi]$ with the forcing
$\vc f(\vc x)=\sin(nx_2)\vc e_1$,
where $\vc e_1=(1,0)^\top$ and $n$ is a positive integer. This choice of forcing yields
the two-dimensional Kolmogorov equation
\begin{subequations}
\beq
\partial_t\vc u =-\vc u\cdot\bnabla\vc u-\bnabla p+\nu \Delta\vc u+\sin(nx_2)\vc e_1,
\eeq
\beq
\bnabla\cdot\vc u=0,
\eeq
\label{eq:kolm}%
\end{subequations}
where, $\vc u=(u_1,u_2)$ and $\nu=1/Re$ with $Re$ being the Reynolds number.

Here, we review some of the relevant properties of this equation that can also be
found in\rf{PlSiFi91} and\rf{CK13}.
The Kolmogorov equation has an equilibrium solution that can be expressed analytically as
\beq
u_1(x_1,x_2)=\frac{Re}{n^2}\sin(nx_2),\quad u_2(x_1,x_2)=0,
\ee{KFeqv0}
at any Reynolds number $Re$. Following~\cite{CK13}, we refer to this solution as the \emph{laminar}
state $E_0$.

The instantaneous energy $E$, energy dissipation $D$ and energy input $I$ of a state $\vc u$ are defined, respectively, by
\begin{subequations}
\begin{alignat}{2}
E(t) &=\frac{1}{2(2\pi)^2}\iint_{\mathbb T^2}|\vc u(\vc x,t)|^2\,\mbox d\vc x,\\
D(t) &=\frac{1}{(2\pi)^2Re}\iint_{\mathbb T^2}|\bnabla\vc u(\vc x,t)|^2\,\mbox d\vc x=
\frac{1}{(2\pi)^2Re}\iint_{\mathbb T^2}|\omega(\vc x,t)|^2\,\mbox d\vc x,\\
I(t)&=\frac{1}{(2\pi)^2}\iint_{\mathbb T^2}u_1(x_1,x_2,t)\sin(nx_2)\,\mbox d\vc x,
\end{alignat}
\label{eq:IDE}%
\end{subequations}
where $|\cdot|$ denotes the usual Euclidean norm and $\omega$ is the non-zero component
of the vorticity field $\bnabla\times\vc u$. One can readily show that
\beq
\frac{\mbox dE}{\mbox d t}=I-D,
\eeq
along an arbitrary solution of the Navier--Stokes equation. For equilibrium and traveling wave
solutions, the energy is time-independent and therefore $I=D=\mbox{const}$.
For the laminar solution $E_0$, for instance, we have
\beq
E_{lam}=\frac{Re^2}{4n^4},\quad D_{lam}=I_{lam}=\frac{Re}{2n^2}.
\eeq

For $n=1$, the laminar state is the
global attractor for all Reynolds numbers,
and therefore all solutions converge to it asymptotically
(see, e.g.,\rf{marchioro86} and \rf{foias2001} (Section 3.1)). For $n>1$ and large
enough
Reynolds numbers, the laminar state becomes unstable. The numerical study of\rf{PlSiFi91}
suggests that for $n=4$ and large enough Reynolds numbers, all invariant solutions of
Kolmogorov equation become unstable, resulting in a chaotic attractor.
More recently,\rf{CK13} confirmed this observation.
To be in agreement with these studies, we also choose the forcing wavenumber $n=4$.
Our numerical results are carried out at $Re=40$, unless otherwise is
mentioned explicitly. \cite{CK13} also carry out their most exhaustive search for invariant
solutions at $Re=40$.

\subsection{Temporal and spatial discretization}
\begin{figure}
\centering
\includegraphics[width=.65\textwidth]{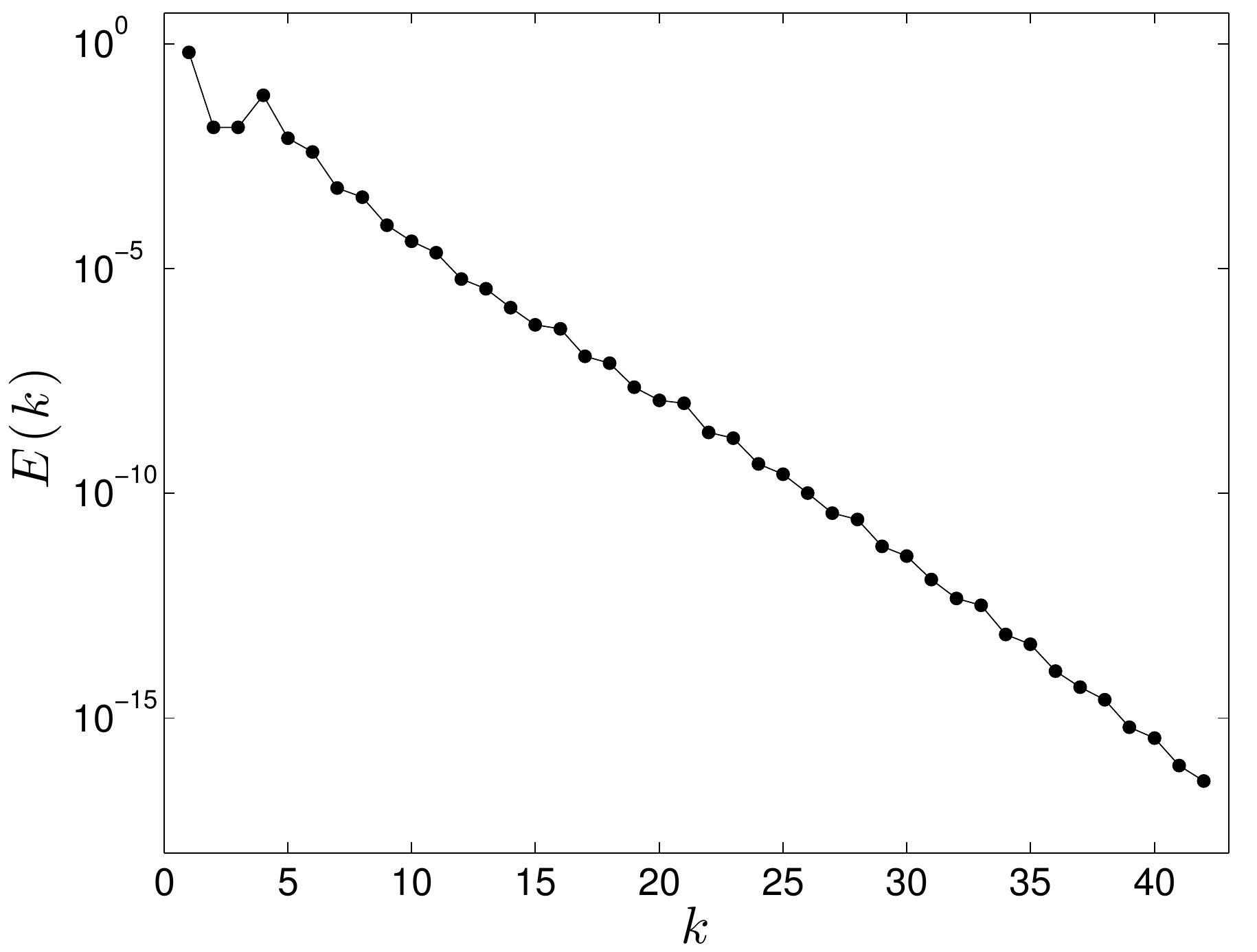}
\caption{Typical energy spectrum of a turbulent trajectory at $Re=40$.}
\label{fig:Ek}
\end{figure}

To evaluate the right-hand-side of the Navier--Stokes equation~\eqref{eq:kolm} and the adjoint
equations~\eqref{eq:adjPDE_NS} and~\eqref{eq:adjPDE_NS_H-1_tw}, we use a standard pseudo-spectral
method with $2/3$ dealiasing. \cite{CK13} consider $Re=40$, $60$, $80$ and $100$ using a uniform
$256\times 256$ Fourier modes for all Reynolds numbers. For $Re=40$, this resolution is
unnecessarily high. Instead, we use $128\times 128$ Fourier modes for $Re=40$.
The resulting energy spectra (cf. Fig.~\ref{fig:Ek}), and the fact that we have been
able to reproduce the (relative) equilibria found by\rf{CK13}, reassures that $128^2$ modes are
sufficient.

For the time integration of the adjoint equations~\eqref{eq:adjPDE_NS}
and~\eqref{eq:adjPDE_NS_H-1_tw},
one can take an explicit adaptive Euler time-step. Denoting the right-hand-side of either adjoint
equation by $\vc G$, the $i$-th Euler step reads
$$\vc u_{i+1}=\vc u_i+\delta\tau_i\,\vc G(\vc u_i),$$
where the length $\delta\tau_i$ of the time step is chosen small enough such that the residue
$\|\vc F(\vc u_{i+1})\|_{\mathcal A}$ is less than $\|\vc F(\vc u_{i})\|_{\mathcal A}$.

\cite{yang07} derive an upper bound for the admissible time step $\delta\tau_i$.
Computing this upper bound for Navier--Stokes equation is, however, not straightforward.
In practice, therefore, one starts with a large value $\delta\tau_i$ and decreases it incrementally
until the residue $\|\vc F(\vc u_{i+1})\|_{\mathcal A}$ reduces compared to the previous iteration.

To gain accuracy, however, we use a higher order numerical scheme for our time integrations.
This is computationally feasible due to
the low cost of evaluating the right-had-sides of equations~\eqref{eq:adjPDE_NS}
and~\eqref{eq:adjPDE_NS_H-1_tw} by the pseudo-spectral method.
More specifically, we use the
embedded Runge--Kutta scheme \texttt{RK5(4)} of\rf{RK45}. This scheme allows for an automatic adaptive time step-size. Roughly speaking,
\texttt{RK5(4)} takes forth and fifth order Runge--Kutta (RK) steps. The fifth order is eventually
used for the time stepping. The forth order prediction is used to adjust the step-size as follows.
Let \texttt{err} denote the absolute difference between the forth and the fifth order predictions.
Then the step-size is chosen such that $\texttt{err}<\texttt{atol}+|\vc u_i|\texttt{rtol}$.
The prescribed constants \texttt{atol} and \texttt{rtol} are the absolute and relative errors,
respectively. We refer the reader to\rf{press:07} (Section 17.2) for further details and an
implementation of the
\texttt{RK5(4)} scheme. This integrator is also implemented in the MATLAB function
\texttt{ode45}.
For integrating the adjoint equations, we choose $\texttt{atol}=\texttt{rtol}=10^{-10}$.
Even with this conservative choice, step sizes as large as $10$ fictitious-time units were
taken by \texttt{RK5(4)}.

We also use \texttt{RK5(4)} for the temporal discretization of the Navier--Stokes
equation~\eqref{eq:kolm}. However, as the time step sizes were much smaller for integrating
this equation, we used the less stringent error tolerances $\texttt{atol}=\texttt{rtol}=10^{-5}$.

\subsection{Hybrid adjoint--Newton iterations}\label{sec:hybrid}
\begin{figure}
\centering
\includegraphics[width=.85\textwidth]{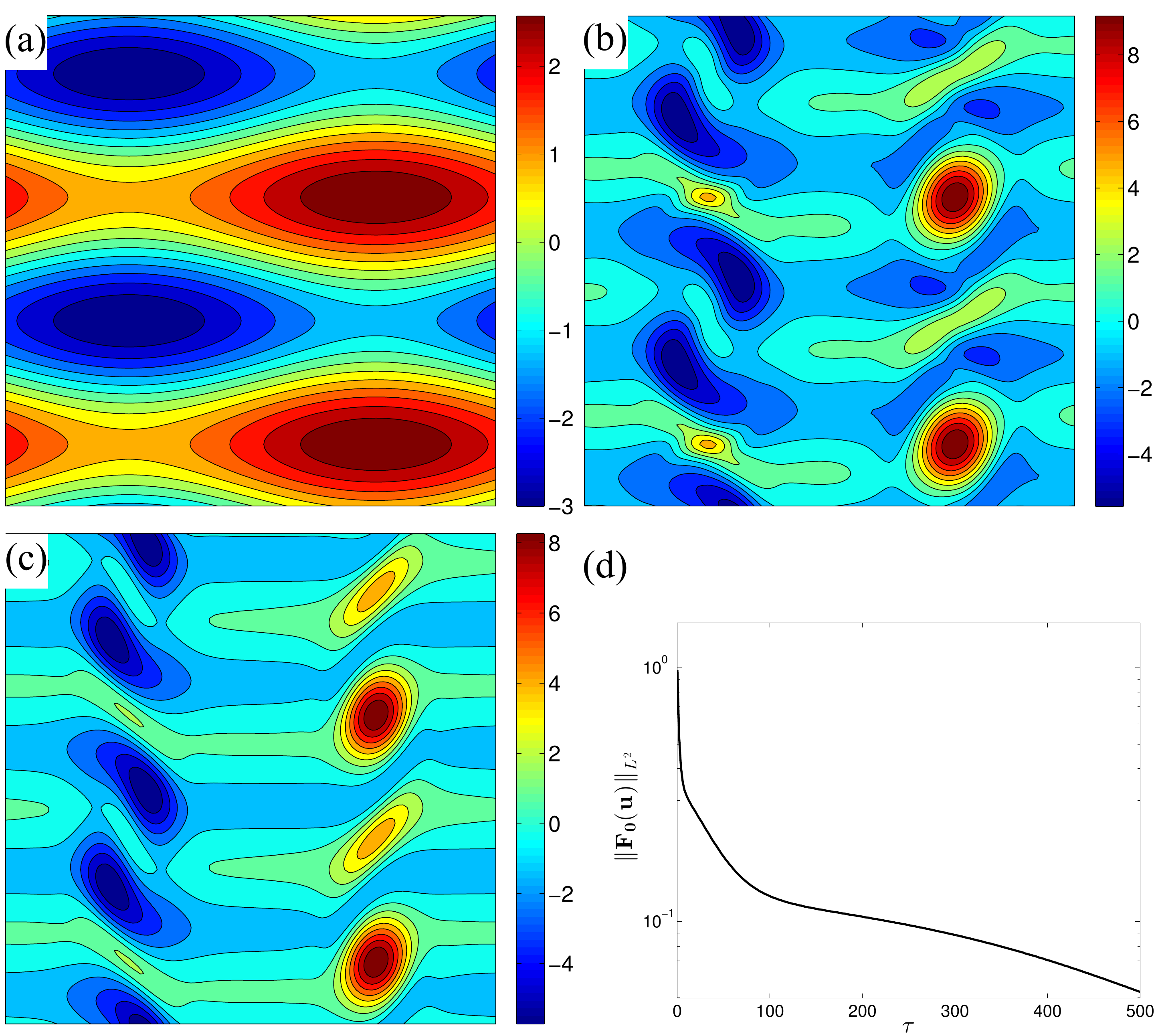}
\caption{
(a) The vorticity field for the initial condition $\vc
u(\vc x,0)=(\cos(2x_2),\cos(x_1))^\top$.
(b) The initial
condition evolved to time $\tau=500$ under the adjoint equation~\eqref{eq:adjPDE_NS} to obtain
$\vc u(\vc x,500)$.
(c) The result $\vc u(\vc x,500)$ shown in panel (b) is used as the initial
guess for the
Newton--GMRES-hook iterations which converged after $7$ iterations with residue
$5.98\times 10^{-11}$. This equilibrium solution is labeled in Table~\ref{tab:EQ} as $E_4$.
(d) The evolution of the $L^2$-residue $\|\vc F_{\vc 0}(\vc u(\cdot,\tau))\|_{L^2}$
as the initial condition in (a) evolves under the adjoint equation from
$\tau=0$ to $\tau=500$}
\label{fig:hybrid}
\end{figure}

We find that the adjoint equation~\eqref{eq:adjPDE_NS} does take arbitrary initial conditions to
Navier--Stokes equilibria. The convergence is, however, rather slow. To demonstrate this, we
take the initial condition $\vc u(\vc x,0)=(\cos(2x_2),\cos(x_1))$ (see Fig.~\ref{fig:hybrid}(a))
and
evolve it under the adjoint equation~\eqref{eq:adjPDE_NS} to time $\tau=500$. The result is
shown in Fig.~\ref{fig:hybrid}(b). This integration took $54$ seconds
(on an iMac with a single processor: Intel Core i5, 2.9 GHz).
The $L^2$ norm of the residue $\vc F_{\vc 0}$, however,
decreases to $\simeq 5\times 10^{-2}$ approximately, indicating that much longer integration times
are required to reduce the residue sufficiently enough, say to $10^{-10}$.

This slow convergence is due to nearly marginal, a priori unknown eigenmodes of the adjoint
operator~\citep{LY07}. For particular wave equations, ad hoc
methods have been proposed to eliminate these modes, and hence speed up the
convergence~\citep{yang07}.
Due to the complexity of the Navier--Stokes equations, it is unclear how such mode elimination
techniques could be employed here.

Using the state $\vc u(\vc x,0)=(\cos(2x_2),\cos(x_1))$ directly as the initial
guess for the NGh iterations does not converge to an equilibrium either: after the first $20$
iterations, the residue plateaued around $2\times 10^{-2}$ and remained so for the $75$ iterations
that were carried out.

Instead, when we use the result of the adjoint integration, i.e. $\vc u(\vc x,\tau=500)$, as the
initial guess for NGh iterations, it converges after seven iterations with residue
$5.98\times 10^{-11}$ (Fig.~\ref{fig:hybrid}(c)). These seven iterations took $156.84$ seconds.

This suggests that, although the convergence of the adjoint equation to the equilibrium is slow, it
takes generic initial guesses to the vicinity of the equilibrium at a relatively low computational
cost. This can be seen by comparing panels (b) and (c) of Fig.~\eqref{fig:hybrid}, which
shows that the state $\vc u(\vc x,\tau=500)$ is quite similar to the equilibrium $E_4$,
and thus likely to lie in its NGh domain of attraction. Indeed, switching to
NGh took this state to the equilibrium solution within a few iterations.

\begin{figure}
\centering
\includegraphics[width=\textwidth]{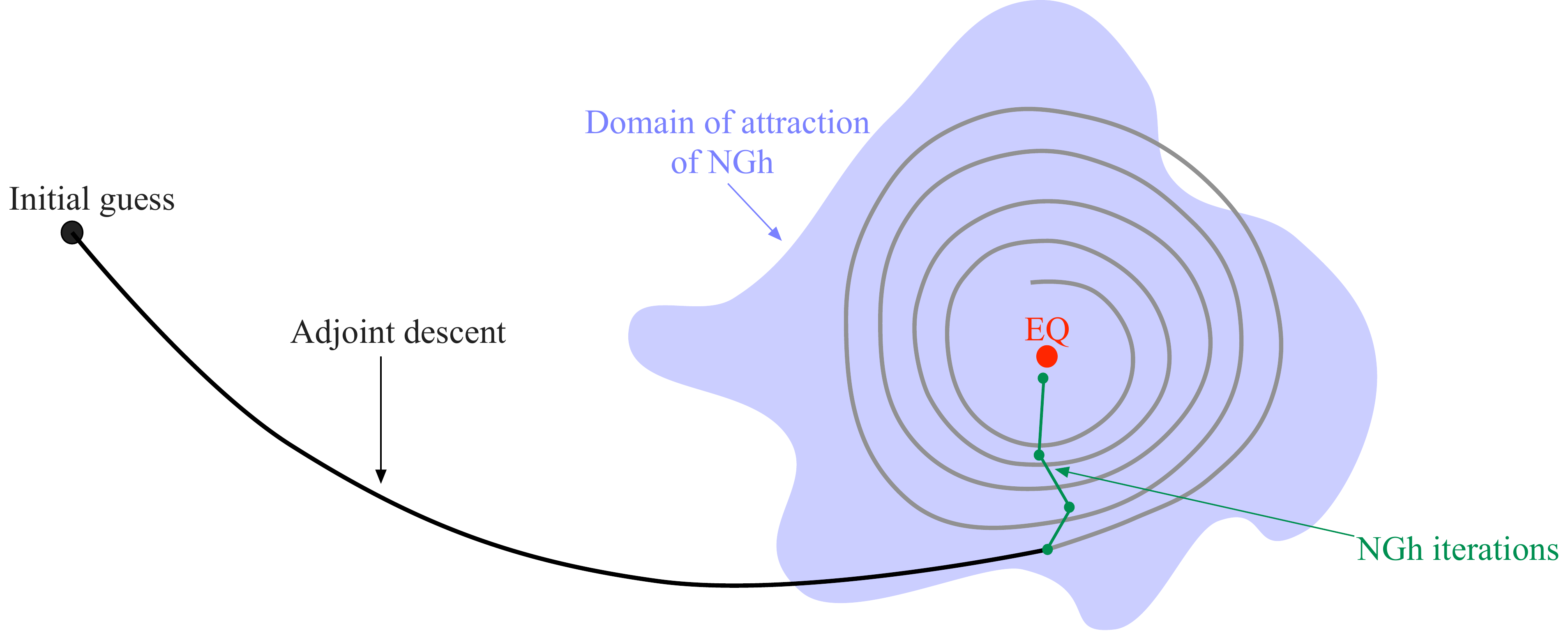}
\caption{
State space depiction of the hybrid adjoint--Newton method. An initial
guess may be too far away from the desired equilibrium solution EQ for
the Newton-GMRES-hook iterations (NGh) to converge to it. The adjoint
descent may eventually converge to EQ, but alone it would take a long
integration time to do so. In the hybrid adjoint--Newton method the
adjoint descent takes the initial guess to the domain of attraction of
NGh. Once there, NGh  converges to EQ in a few iterations.
}
\label{fig:hybrid_schem}
\end{figure}

Based on this observation, we propose the following hybrid adjoint-Newton iterations,
sketched in Fig.~\ref{fig:hybrid_schem}.
We take
a prescribed positive real number $\tau_0$ and a positive integer \texttt{N}. An initial
condition $\vc u(\vc x,0)$ is then integrated under the adjoint equation~\eqref{eq:adjPDE_NS} to
obtain $\vc u(\vc x,\tau_0)$. This state is then fed into the NGh algorithm and \texttt{N}
iterations of NGh are performed. The output is then fed back into the adjoint equation
as the initial condition. This hybrid loop is repeated until the residue $\|\vc F_{\vc 0}\|_{L^2}$
decreases below some prescribed tolerance \texttt{tol}. This procedure is summarized
in Algorithm~\ref{alg:hybrid_eq}.
The hybrid adjoint-Newton iterations for finding traveling waves are similar except
that the adjoint equations~\eqref{eq:adjPDE_NS_H-1_tw} are solved at each iteration
(see Algorithm~\ref{alg:hybrid_tw}).
\begin{algorithm}
\DontPrintSemicolon
\caption{Hybrid adjoint-Newton algorithm for finding equilibrium solutions of the
Navier--Stokes equation.}
\textbf{Input:} $\tau_0,\texttt{tol}\in\mathbb R^+$,
\texttt{N}$\,\in\mathbb N$, state $\vc
u_0$\;
\While{\textnormal{$\|\vc F_{\vc 0}(\vc u_0)\|_{L^2}>\texttt{tol}$}}{
Integrate the adjoint equation~\eqref{eq:adjPDE_NS} for $\tau_0$ fictitious-time units
with the initial condition $\vc u(\cdot,0)=\vc u_0$\;
$\vc u_0\longleftarrow\vc u(\cdot,\tau_0)$\;
\If{\textnormal{$\|\vc F_{\vc 0}(\vc u_0)\|_{L^2}<\texttt{tol}$}}{STOP}
\For{\textnormal{($k=1$ \KwTo \texttt{N})}}{
Take one Newton--GMRES-hook step to get NGh($\vc u_0$)\;
$\vc u_0\longleftarrow$ NGh($\vc u_0$)\;
\If{\textnormal{$\|\vc F_{\vc 0}(\vc u_0)\|_{L^2}<\texttt{tol}$}}{STOP}
}
}
\textbf{Output:} $\vc u_0$
\label{alg:hybrid_eq}
\end{algorithm}
\begin{algorithm}
\DontPrintSemicolon
\caption{Hybrid adjoint-Newton algorithm for finding traveling wave solutions of the
Navier--Stokes equation.}
\textbf{Input:} $\tau_0,\texttt{tol}\in\mathbb R^+$,
\texttt{N}$\,\in\mathbb N$, state $\vc u_0$, wave speed $\vc c_0$\;
\While{\textnormal{$\|\vc F_{\vc c_0}(\vc u_0)\|_{L^2}>\texttt{tol}$}}{
Integrate the adjoint equation~\eqref{eq:adjPDE_NS_H-1_tw} for $\tau_0$ fictitious-time units
with the initial conditions $\vc u(\cdot,0)=\vc u_0$ and $\vc c(0)=\vc c_0$\;
$\vc u_0\longleftarrow\vc u(\cdot,\tau_0)$\;
$\vc c_0\longleftarrow\vc c(\tau_0)$\;
\If{\textnormal{$\|\vc F_{\vc c_0}(\vc u_0)\|_{L^2}<\texttt{tol}$}}{STOP}
\For{\textnormal{($k=1$ \KwTo \texttt{N})}}{
Take one Newton--GMRES-hook step to get NGh($\vc u_0,\vc c_0$)\;
$(\vc u_0,\vc c_0)\longleftarrow$ NGh($\vc u_0,\vc c_0$)\;
\If{\textnormal{$\|\vc F_{\vc c_0}(\vc u_0)\|_{L^2}<\texttt{tol}$}}{STOP}
}
}
\textbf{Output:} $\vc u_0, \vc c_0$
\label{alg:hybrid_tw}
\end{algorithm}

For the computations reported below, we set $\tau_0=100$, $\texttt{tol}=10^{-10}$ and
$\texttt{N}=1$. Since NGh steps are relatively expensive, we only take one NGh step
($\texttt{N}=1$) per iteration of
adjoint-Newton.
At Reynolds number $Re=40$, for instance, the integration of the adjoint equations
to $\tau=100$
took approximately $10$ seconds on average while each NGh step took approximately $50$ seconds
on average.

\subsection{Equilibrium solutions}
In this section, we report the equilibria found by the hybrid adjoint-Newton iterations
(Algorithm~\ref{alg:hybrid_eq}).
We also compare its performance with that of pure NGh iterations without the
adjoint step.

To this end, we consider the set of initial guesses
\beq
\vc u_0^{(m_1,m_2)}=\big(\cos(m_2x_2),\cos(m_1x_1)\big),
\label{eq:ig}
\eeq
for a range of integers $m_1$ and $m_2$. We find that for $m_2>4$, both Newton--GMRES-hook
and the hybrid adjoint-Newton iterations converge to the laminar equilibrium $E_0$. Therefore, we
restrict the range of the integers to $1\leq m_1,m_2\leq 4$. This leads to $16$ distinct
initial guesses.
\begin{figure}
\centering
\includegraphics[width=\textwidth]{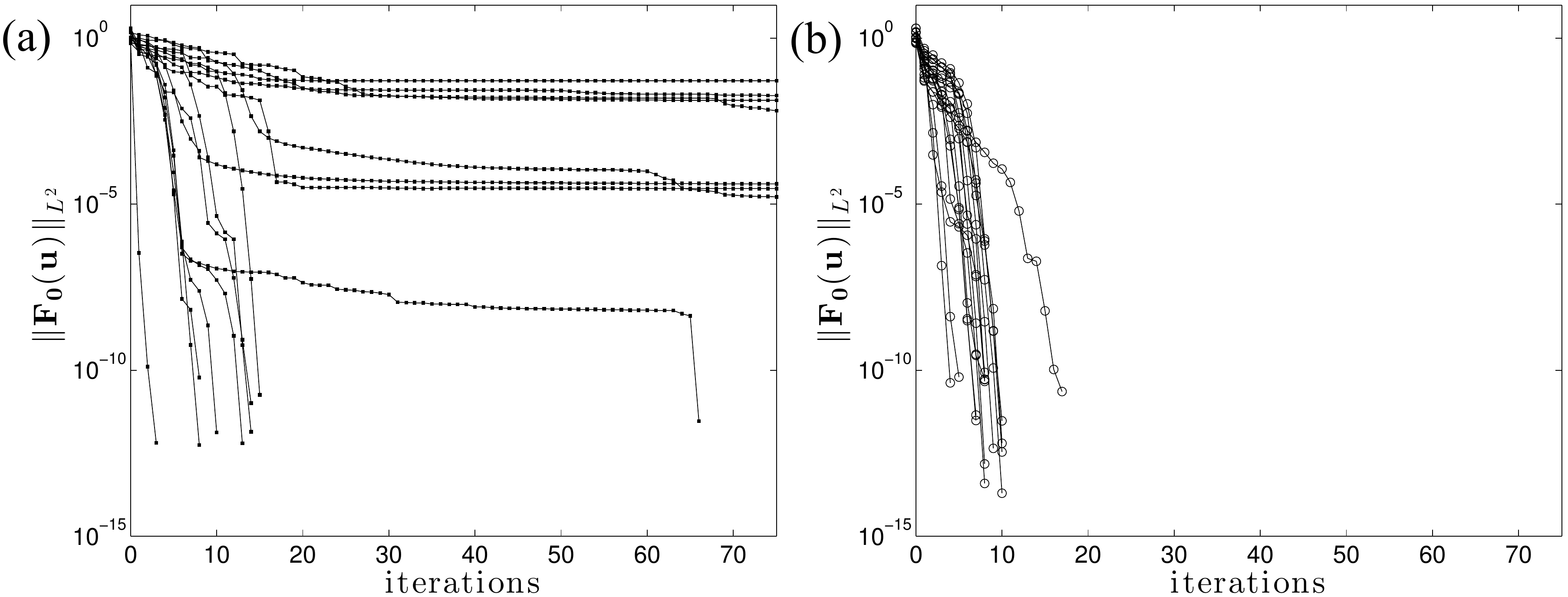}
\caption{The decay of the residue (i.e., the $L^2$-norm of the right hand side) using
Newton-GMRES-hook iterations (a) and the hybrid
adjoint-Newton iterations (b).}
\label{fig:NGh_vs_adjN}
\end{figure}

\begin{figure}
\centering
\includegraphics[width=\textwidth]{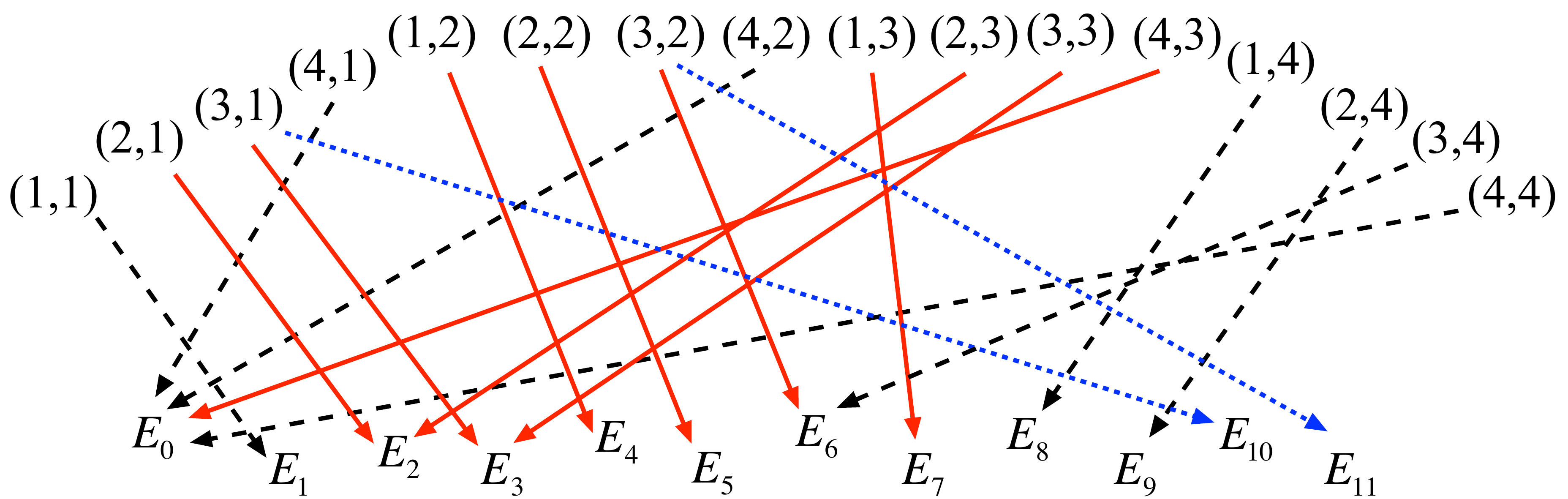}
\caption{
The convergence diagram. Double indices $(m_1,m_2)$ label the initial guesses
\eqref{eq:ig}.
A dashed black arrow indicates that both the hybrid adjoint-Newton and the NGh
iterations converge to the same equilibrium.
A dotted blue arrow indicates that the NGh iterations converge to a different
equilibrium from the one reached by the hybrid adjoint-Newton (red
arrow).
A solid red arrow (with no blue arrow originating from the same initial guess)
indicates that only the hybrid adjoint-Newton converged to the equilibrium,
i.e., the NGh iterations alone failed to converge.
}
\label{fig:EQ_diag}
\end{figure}

These initial conditions are divergence-free by construction, consistent with our incompressible
Navier--Stokes setting. They are also explicit, rendering the following results
reproducible. Furthermore, the generic nature of the initial conditions illustrates the main
advantage of the adjoint descent method, namely that its convergence does not require
good initial guesses.
For brevity, we shall refer to the states $\vc u_0^{(m_1,m_2)}$ by their
indices $(m_1,m_2)$.

\begin{table}
\centering
\begin{tabular}{c c c c c c}
Equilibrium &  I=D       & E          & $\mu_1$ & $\omega_1$ &  $\dim E^u$ \\
\hline\\
$E_0$    &  1.25000   & 1.56250    & 2.35340 &   0.0      & 38  \\
$E_1$    &  0.12732   & 0.76108    & 0.24929 &  2.56010   & 9   \\
$E_2$    &  0.15406   & 0.47313    & 0.66722 &   0.0      & 13  \\
$E_3$    &  0.26287   & 0.45173    & 0.84744 &   0.0      & 27  \\
$E_4$    &  0.08433   & 0.57317    & 0.62697 &   0.0      & 5   \\
$E_5$    &  0.26661   & 0.55183    & 0.30119 &   0.0      & 19  \\
$E_6$    &  0.26227   & 0.45574    & 0.86924 &   0.0      & 22  \\
$E_7$    &  0.07530   & 0.58104    & 0.58318 &   0.0      & 5   \\
$E_8$    &  0.17452   & 0.53493    & 0.61189 &   0.00675  & 17  \\
$E_9$    &  0.15315   & 0.47396    & 0.67006 &   0.0      & 17  \\
$E_{10}$ &  0.61437   & 0.91740    & 0.85655 &   1.75030  & 30  \\
$E_{11}$ &  0.48020   & 0.71284    & 0.81175 &   0.0      & 29  \\
$E_{12}$ &  0.31049   & 0.89032    & 0.68928 &   0.0      & 13  \\
$E_{13}$ &  0.26151   & 0.50921    & 0.42190 &   1.49660  & 16  \\
$E_{14}$ &  0.31152   & 0.54066    & 0.71722 &   0.0      & 21  \\
$E_{15}$ &  0.27070   & 0.76570    & 0.87760 &   0.0      & 14  \\
$E_{16}$ &  0.34954   & 0.61168    & 0.83092 &   0.02382  & 24  \\
\end{tabular}
\caption{List of equilibrium solutions at $Re=40$.
Energy $E$, energy dissipation $D$ and energy input $I$ are
	defined in \eqref{eq:IDE}.
    The leading unstable
	eigenvalue of the equilibrium is $\mu_1+i\omega_1$. The dimension of the linear unstable
    manifold of the equilibrium is denoted by $\dim E^u$.}
\label{tab:EQ}
\end{table}

Using the hybrid adjoint-Newton iterations for equilibria (see Algorithm~\ref{alg:hybrid_eq}), all
$16$ runs converged, resulting in $10$ distinct equilibria. These equilibria are listed as $E_0$
to $E_9$ in Table~\ref{tab:EQ}, where $E_0$ is the laminar state~\eqref{KFeqv0}.
All initial guesses converged to an equilibrium within the first $10$ iterations
of hybrid adjoint-Newton iterations, except initial guess $(2,1)$ that took $17$ iterations
(see Fig.~\ref{fig:NGh_vs_adjN}). Fig.~\ref{fig:EQ_diag} shows the convergence diagram, connecting
each initial guess $(m_1,m_2)$ to the resulting equilibrium solution.

Using NGh alone, only $9$ out of $16$ runs converged. They converged to $6$ distinct
equilibria: $E_0$, $E_1$, $E_6$, $E_8$, $E_9$, $E_{10}$ and $E_{11}$
(see Fig.~\ref{fig:EQ_diag}).
This comes initially as a surprise since the exhaustive search carried out by\rf{CK13} only
returned
a single equilibrium (i.e., $E_1$).
This can be explained, however, in terms of the method used for initiating
the NGh iterations. \cite{CK13} use recurrences to find initial guesses for their searches, as
opposed to the generic initial guesses used here. Recurrences only happen in a subset of the
state space where a generic turbulent trajectory spends most of its lifetime.
As a result, using recurrences for initiating the NGh searches
might preferentially yield the equilibria `close' to this region.
We will return to this subject in Section~\ref{sec:interm}, where the temporal
intermittency is discussed.
\begin{figure}
\includegraphics[width=.32\textwidth]{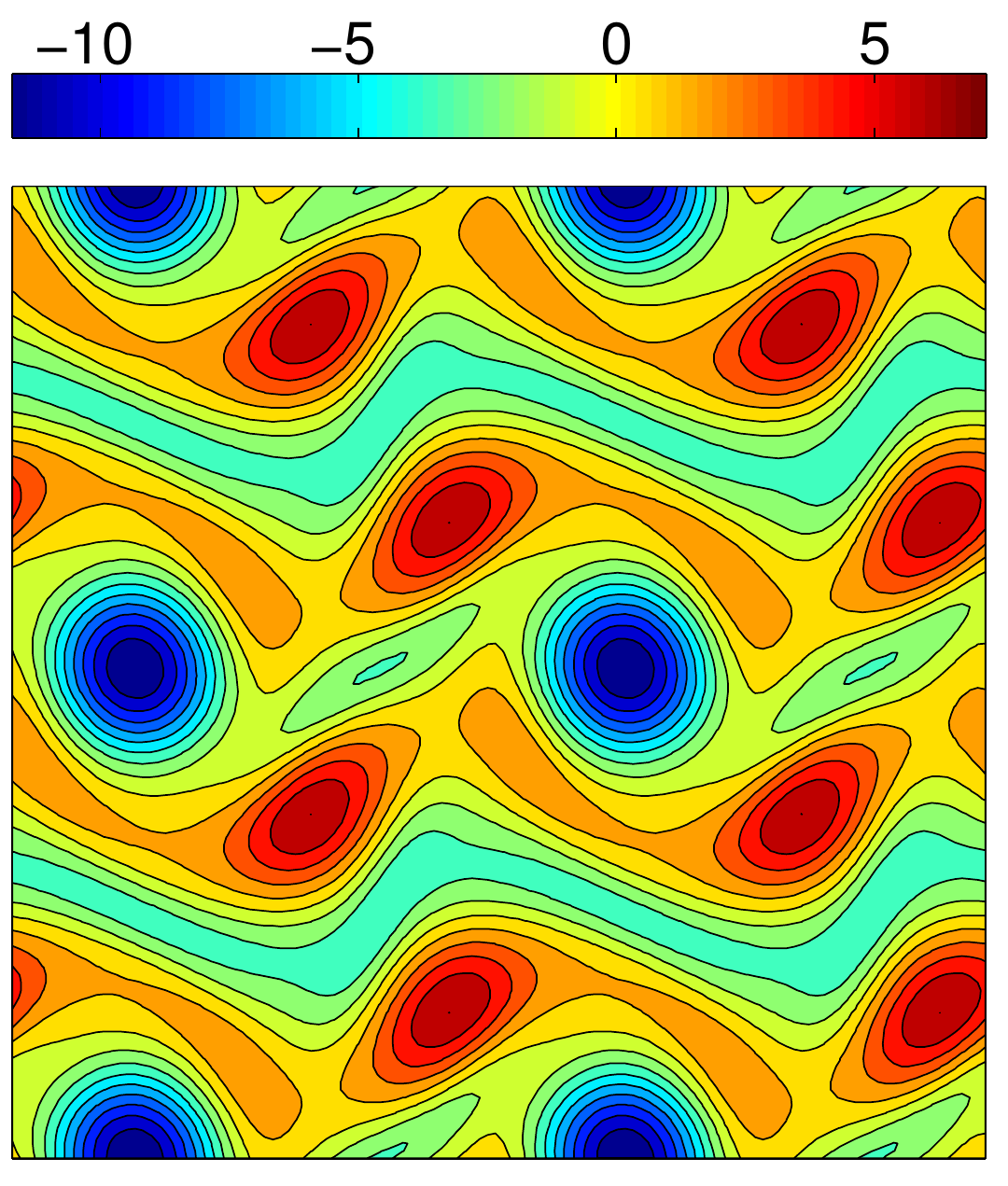}
\includegraphics[width=.32\textwidth]{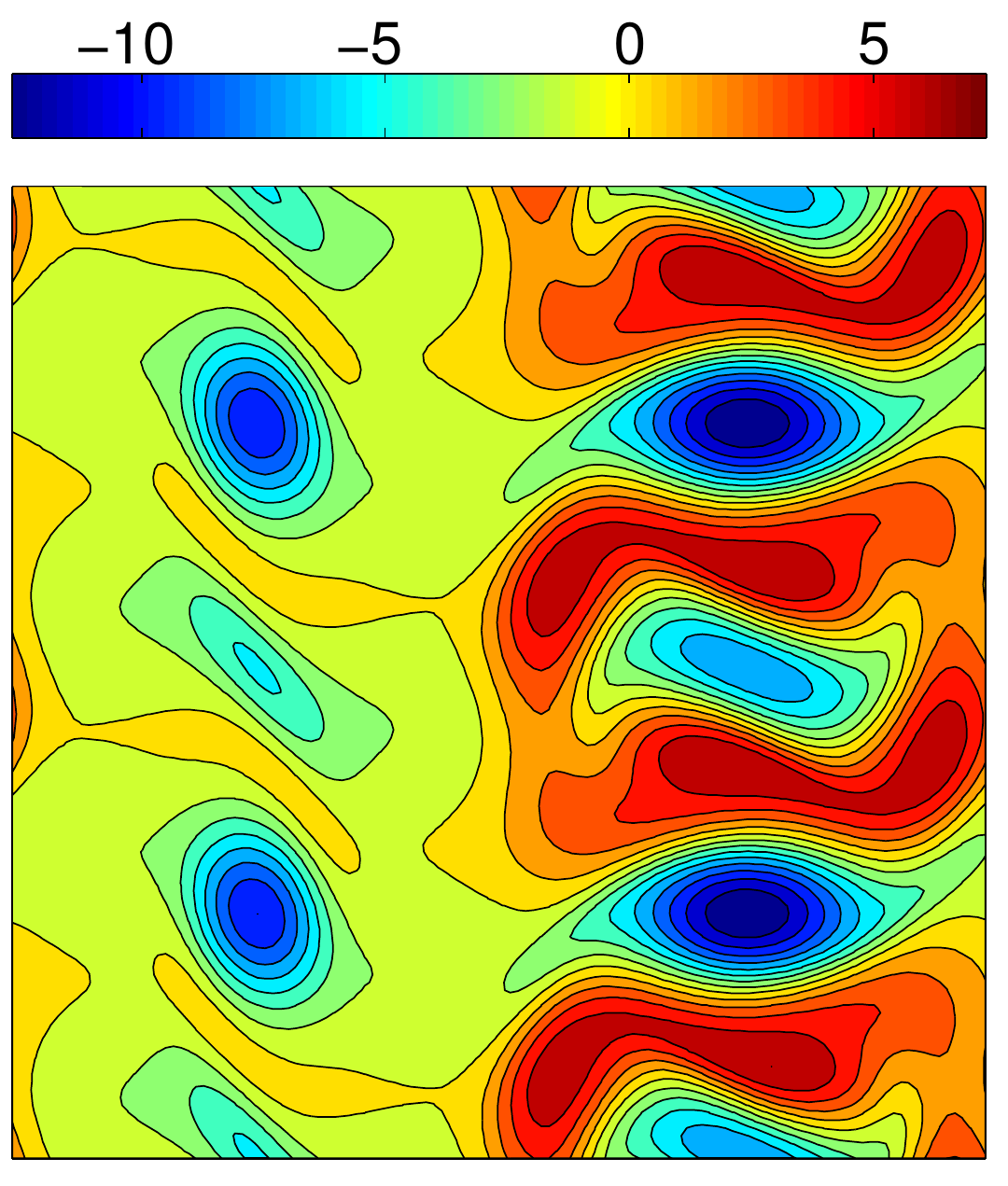}
\includegraphics[width=.32\textwidth]{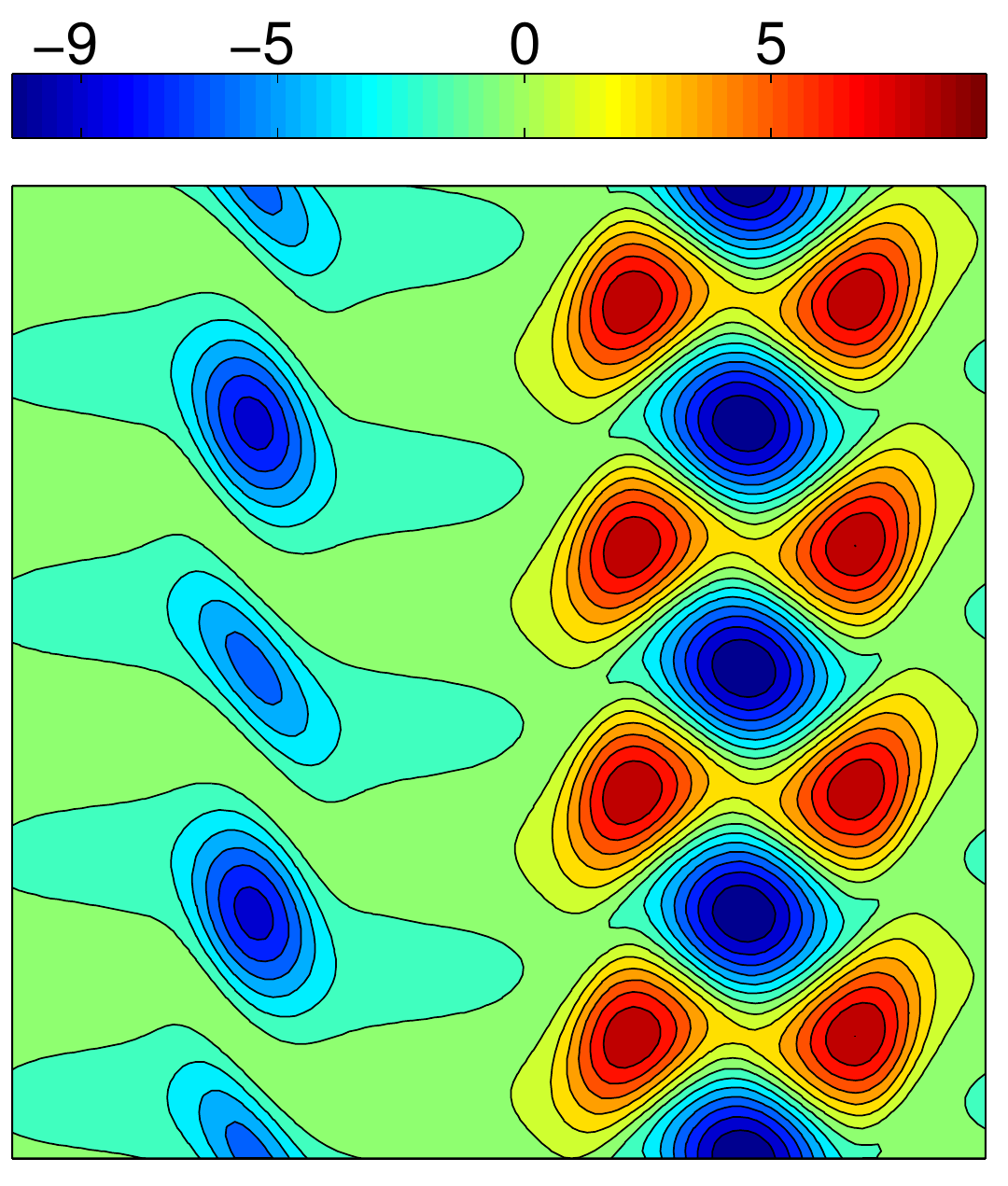}
\caption{Vorticity fields of the equilibrium solutions $E_5$ (left), $E_{12}$ (middle)
and $E_{15}$ (right). All panels show the entire domain $[0,2\pi]\times[0,2\pi]$}
\label{fig:EQ}
\end{figure}

We also searched for further equilibria using states of the form
$(\sin(m_2x_2),\cos(m_1x_1))$ as the initial guesses for the hybrid adjoint-Newton iterations.
This resulted into five more equilibria: $E_{12}$ to $E_{16}$ in Table~\ref{tab:EQ}.
While all of these additional searches did converge,
most of them re-converged to previously discovered
equilibria, including $E_{10}$ and $E_{11}$ that were only found by NGh iterations
when initial guesses~\eqref{eq:ig} were used. Figure~\ref{fig:EQ} shows the
vorticity field of three select equilibrium solutions.
\begin{figure}
\includegraphics[width=.32\textwidth]{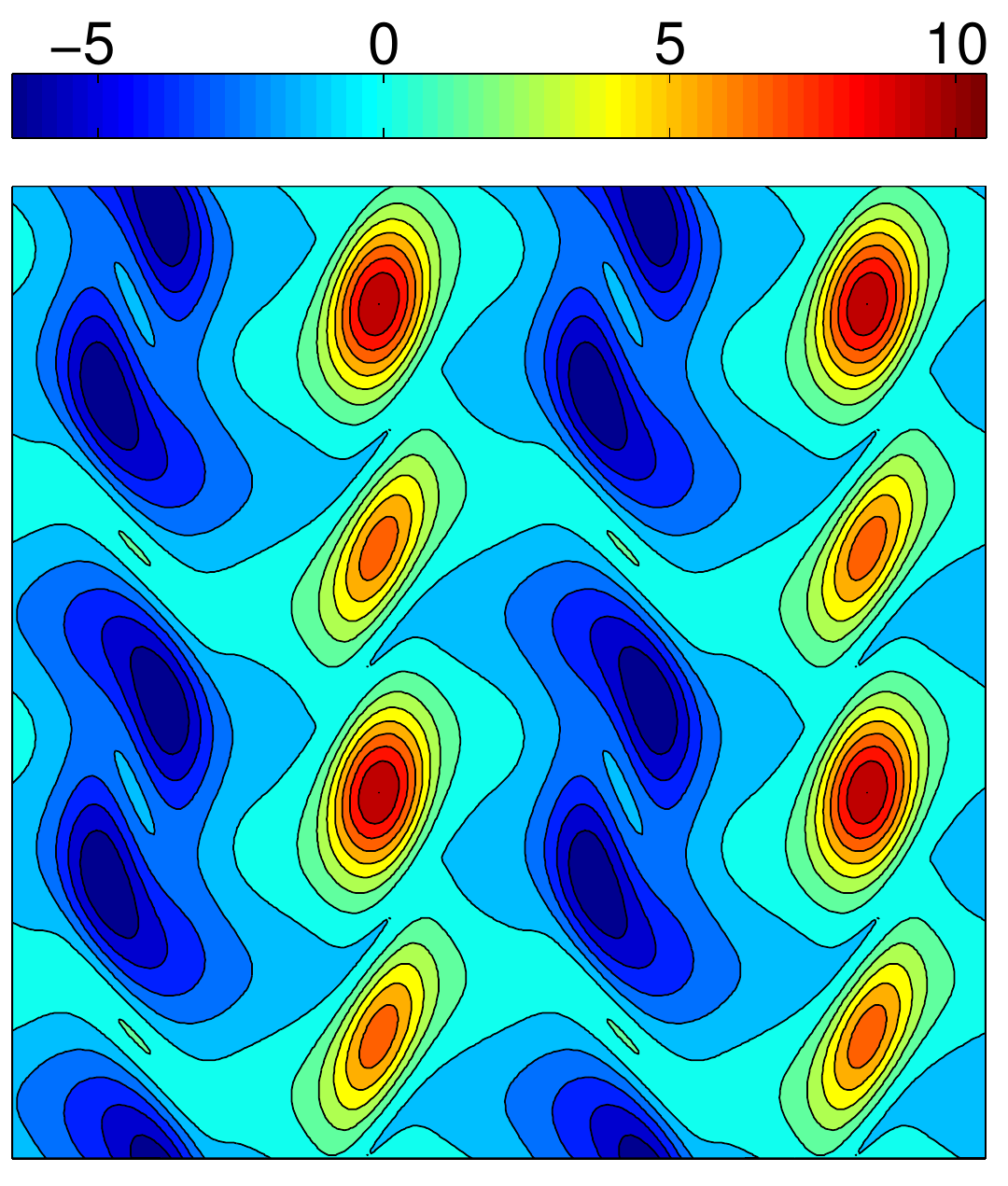}
\includegraphics[width=.32\textwidth]{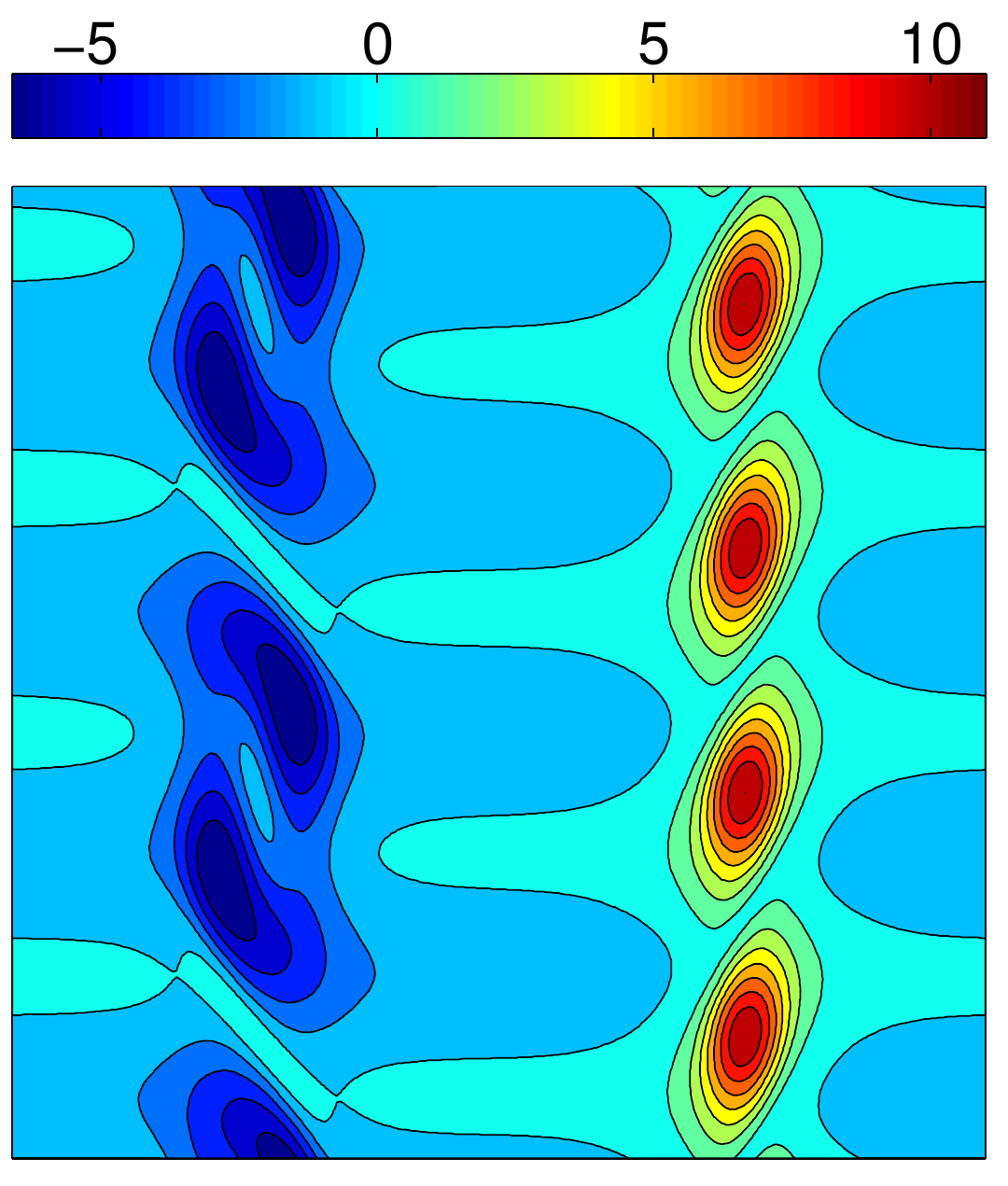}
\includegraphics[width=.32\textwidth]{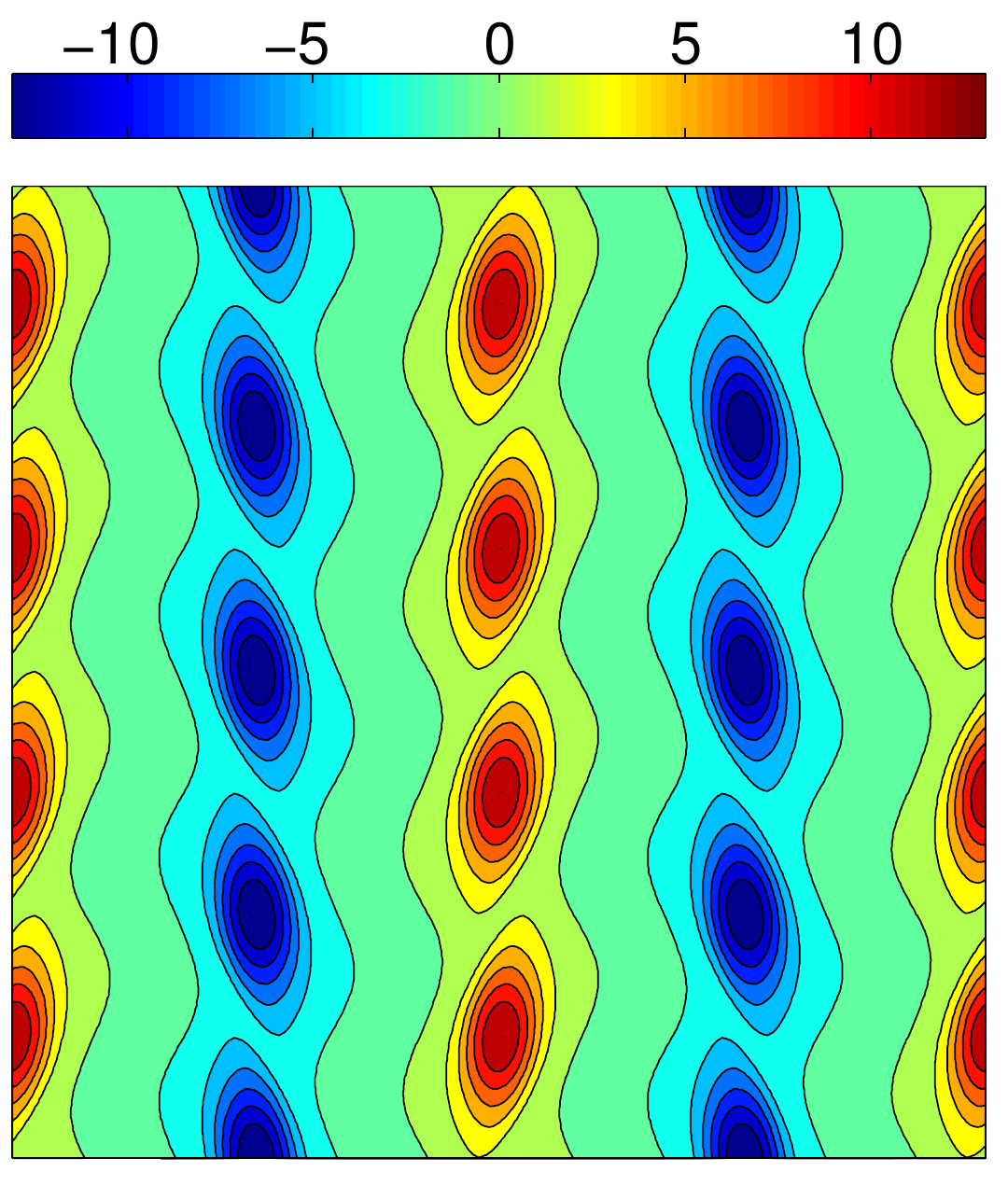}
\caption{Vorticity fields of the equilibrium solutions at Reynolds number
$Re=60$ (left), $Re=80$ (middle) and $Re=100$
(right). All panels show the entire $[0,2\pi]\times[0,2\pi]$ domain.
}
\label{fig:EQ_high}
\end{figure}

Some of the equilibria (e.g., $E_{15}$ shown in Fig.~\ref{fig:EQ}) exhibit vertical bands
of localized vorticity that are separated by an almost zero vorticity background.
Such spatially localized equilibria of
the Kolmogorov flow were only observed previously on domains with small aspect ratio
$\alpha=L_2/L_1$~\citep{LK14}. The fact that they also exist on a domain with
aspect ratio $\alpha=1$ comes as a surprise.

Although our focus here is on Reynolds number $Re=40$,
Fig.~\ref{fig:EQ_high} showcases select equilibria
at $Re=60$, $80$ and $100$.
These equilibria are computed using the higher resolution of $256\times 256$ Fourier modes.
They are found by our hybrid adjoint-Newton method while
the previous studies using Newton--GMRES-hook iterations had not been
able to discover them~\citep{CK13}.

\subsection{Traveling wave solutions}
The forcing term in the Kolmogorov equation~\eqref{eq:kolm}
breaks the continuous symmetry in the $x_2$-direction. Therefore, only traveling wave
solutions of the type $\vc u(\vc x-\vc ct)$ with $\vc c=(c,0)$ are permitted.
This reduces the wave velocity equations~\eqref{eq:dot_c} to the scalar equation
\beq
\frac{\mathrm d c}{\mathrm d \tau}=-\int_{\mathbb T^2}
\frac{\partial\vc u}{\partial x_1}\cdot\tilde{\vc
u}''\;\mathrm{d}^2\vc x.
\eeq
Similarly, the term $\vc c\cdot\bnabla\tilde{\vc u}''$ in~\eqref{eq:adjPDE_tw-1} reduces to
$c\,\partial_{x_1}\tilde{\vc u}''$ and the term $\vc c\cdot\bnabla\vc u$
in~\eqref{eq:u''_tw} reduces to $c\,\partial_{x_1}\vc u$.

We search for traveling waves using Algorithm~\ref{alg:hybrid_tw} and generic
initial guesses discussed in the previous section. For the initial
wave speed, we used $c(0)=1$. Other values of $c(0)$ yielded similar results.
\begin{table}
\centering
\begin{tabular}{c c c c c c c}
Traveling wave & c &  I=D & E & $\mu_1$ & $\omega_1$ &  $\dim E^u$ \\
\hline\\
$T_1$ & 0.01978 & 0.08873 & 0.69747 & 0.06815 & 0.35451 & 4 \\
$T_2$ & 0.00944 & 0.08680 & 0.63969 & 0.45288 & 0.02134 & 4 \\
$T_3$ & 0.01826 & 0.13432 & 0.38056 & 0.49301 & 0.0 & 10 \\
$T_4$ & 0.03062 & 0.31453 & 0.52793 & 0.55183 & 0.0 & 21 \\
$T_5$ & 0.05266 & 0.40027 & 0.61651 & 0.82112 & 0.0 & 17 \\
$T_6$ & 0.04223 & 0.32063 & 0.52963 & 0.54292 & 0.0 & 18 \\
$T_7$ & 0.00208 & 0.08058 & 0.61200 & 0.50598 & 0.0 & 2 \\
$T_8$ & 0.00156 & 0.09641 & 0.59883 & 0.63357 & 0.0 & 7 \\
$T_9$ & 0.00642 & 0.08867 & 0.61764 & 0.57681 & 0.0 & 4 \\
\end{tabular}
\caption{List of traveling wave solutions at $Re=40$. The constant $c$ denotes the
wave speed. Energy $E$, energy dissipation $D$ and energy input $I$ are
defined in \eqref{eq:IDE}. The leading unstable
eigenvalue of the traveling wave is $\mu_1+i\omega_1$.
The dimension of the linear unstable manifold of
the traveling wave is denoted by $\dim E^u$.}
\label{tab:TW}
\end{table}
\begin{figure}
\includegraphics[width=.32\textwidth]{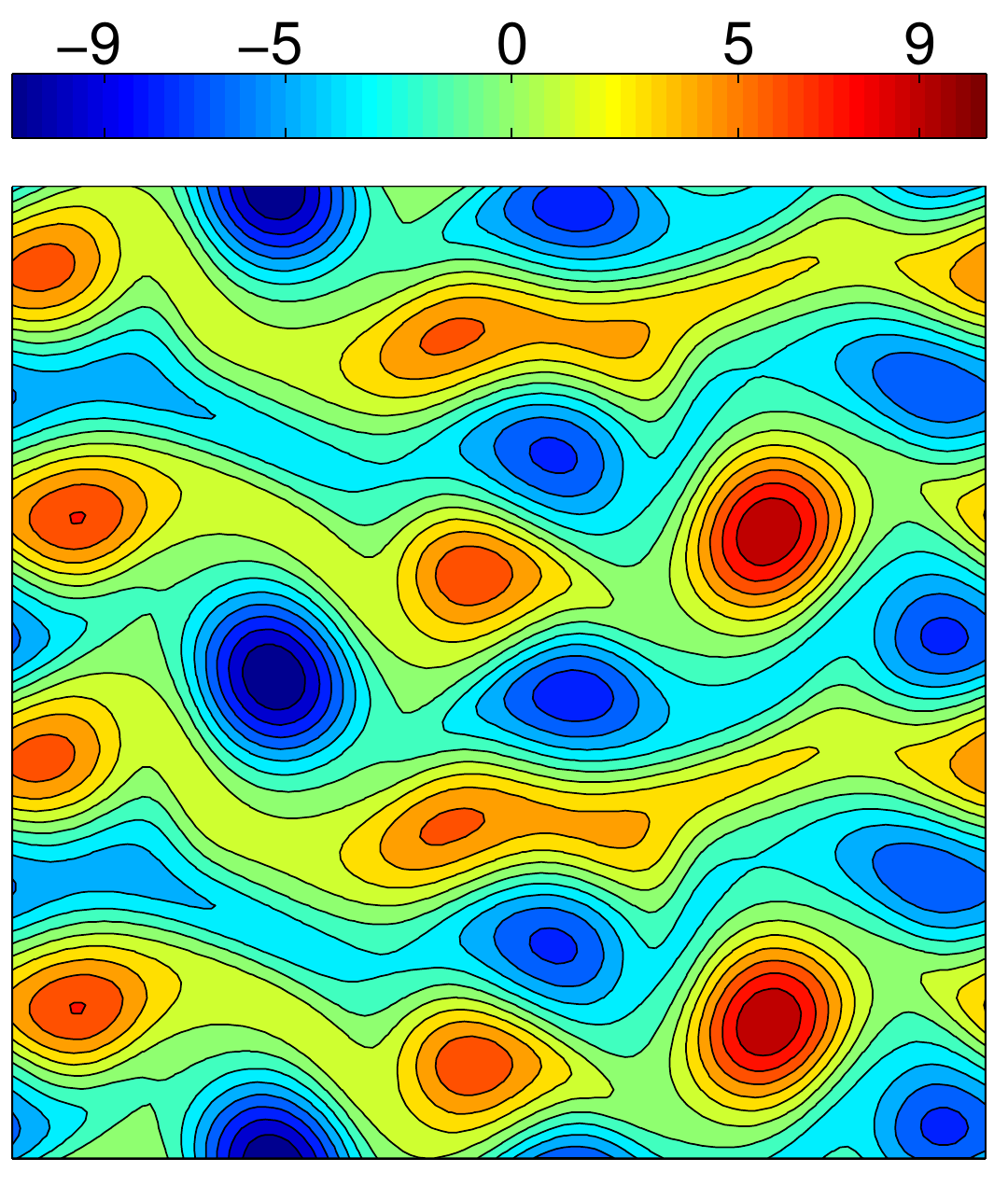}
\includegraphics[width=.32\textwidth]{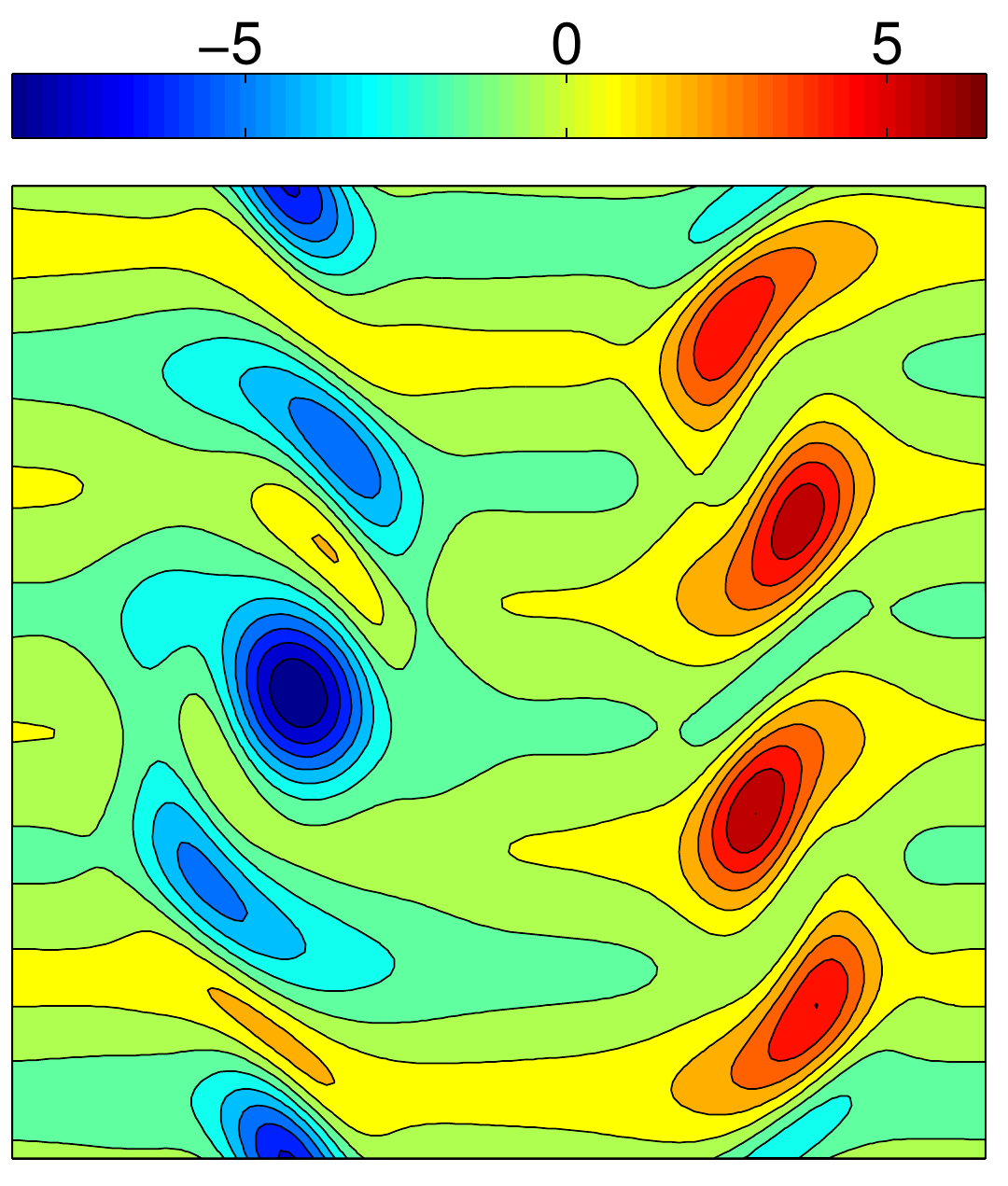}
\includegraphics[width=.32\textwidth]{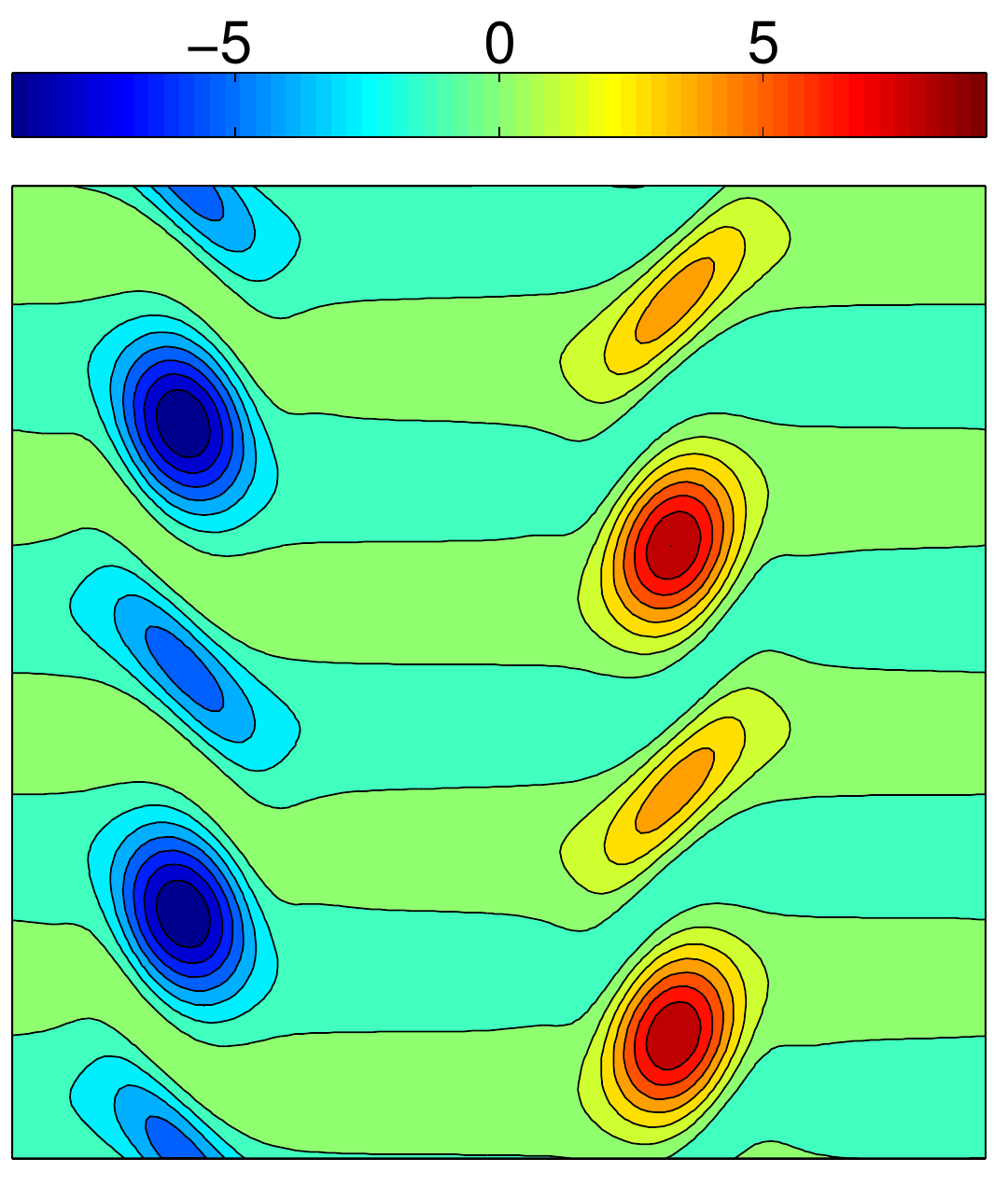}
\caption{Vorticity field for the traveling wave solutions $T_4$ (left), $T_{7}$ (middle) and
$T_{8}$ (right). All panels show the entire domain $[0,2\pi]\times[0,2\pi]$.}
\label{fig:TW}
\end{figure}

Some of our searches for traveling waves, converged to equilibrium solutions.
This is to be expected as equilibria are degenerate traveling
waves with wave speed $c=0$.
In fact, the adjoint equation~\eqref{eq:adjPDE_NS_H-1_tw} admits such solutions. This can
be readily verified by letting $\vc u(\vc x)$ to be an equilibrium solution of the Navier--Stokes
equation and setting $\vc c\equiv 0$. Then $\vc u''$ in~\eqref{eq:u''_tw} is zero, resulting
in vanishing right-hand-sides in equations~\eqref{eq:adjPDE_NS_H-1_tw}.

Nonetheless, our hybrid adjoint-Newton searches led to $9$ distinct traveling waves listed
in Table~\ref{tab:TW}. Only traveling waves $T_1$ and $T_2$ had been discovered
previously~\citep{CK13}.
Figure~\ref{fig:TW} shows the vorticity field for three select traveling waves.
As in the case of equilibria, we find that some of the traveling waves (e.g., $T_8$ in
Fig.~\ref{fig:TW}) exhibit localized spatial structures, although the domain aspect ration is one.

\section{Temporal intermittency in Kolmogorov flow}\label{sec:interm}
In this section, we explore the significance of the invariant solutions, 
found in the previous section,
on the global dynamics of the Kolmogorov flow.

\begin{figure}
\centering
\includegraphics[width=.75\textwidth]{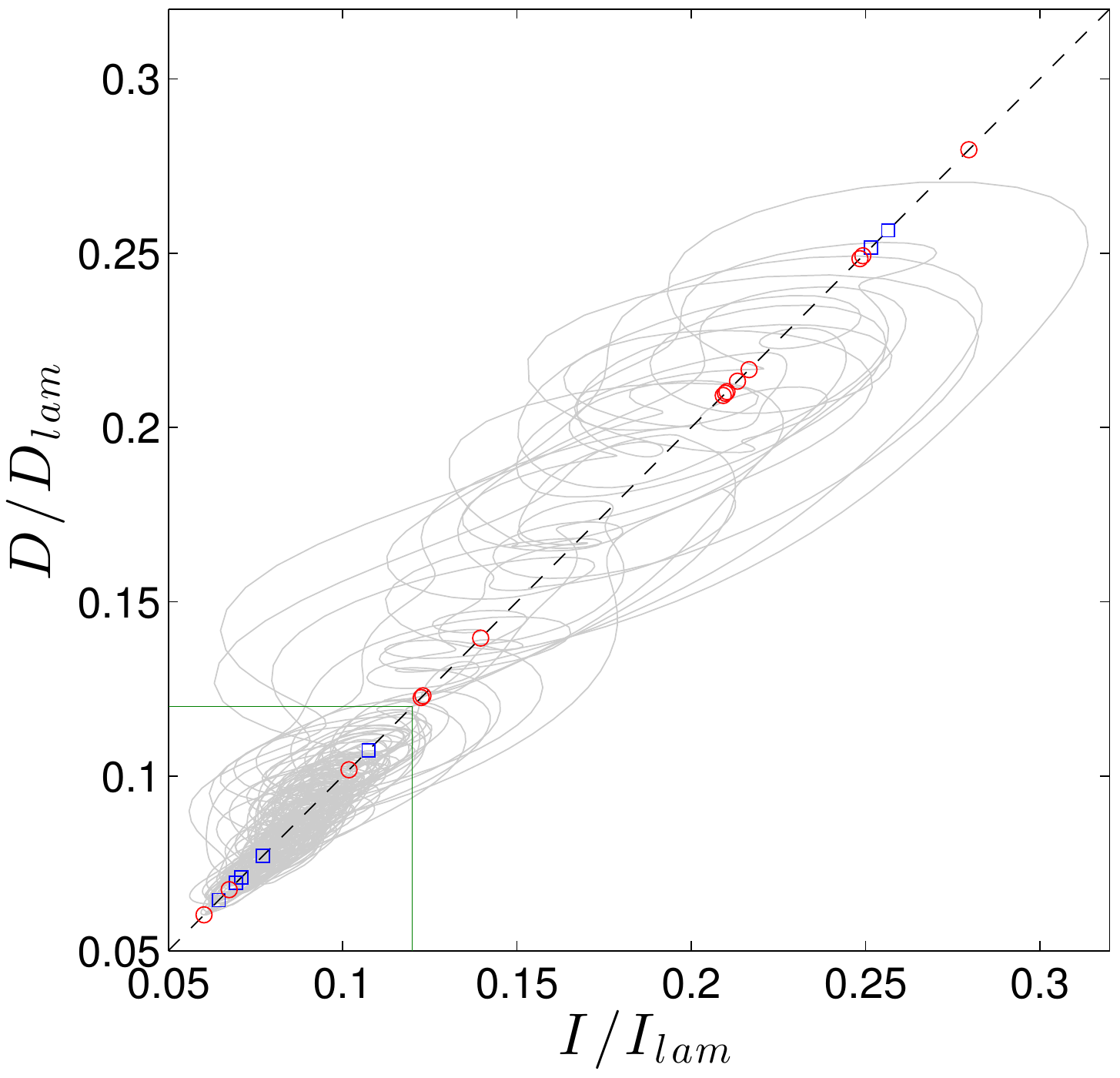}		
\caption{$\mbox{Re}=40$. Gray: Turbulent trajectory spanning $10^3$ time units.
Red circles: equilibria.
Blue squares: traveling waves.
The green square marks the region where $I/I_{lam}<0.12$ and $D/D_{lam}<0.12$.
The turbulent trajectory spends $86.62\%$ out of the total $10^3$ time units inside this region.
The diagonal $I=D$ is marked by the dashed black line.
Equilibria and traveling waves with $I/I_{lam}=D/D_{lam}>0.32$ are not shown.
}
\label{fig:ID_R40}
\end{figure}

Figure~\ref{fig:ID_R40} shows the energy input $I$ versus the energy dissipation $D$
for a generic turbulent trajectory computed for $10^3$ time units
and recorded every $0.1$ time units.
The energy input
and dissipation of each equilibrium and traveling wave are marked by circles and squares,
respectively. As mentioned earlier, the energy input and dissipation coincide for these invariant
solutions, locating them on the diagonal $I=D$.

The equilibria and traveling waves assume a wide range of energy input and dissipation.
The turbulent trajectory, on the other hand,
mostly resides in a relatively low energy input/dissipation
regime. In particular, approximately $85\%$ of this trajectory belongs to the $I/I_{lam}<0.12$ and
$D/D_{lam}<0.12$ regime, marked by the green square in Fig.~\ref{fig:ID_R40}. For the lack
of a better term, we refer to this regime as the `ergodic sea'.

At the same time, the turbulent trajectory also experiences sporadic episodes of high energy input
and dissipation. Such rare, extreme events are usually referred to as \emph{temporal intermittency}
and are ubiquitous in turbulent fluid flow (see, e.g.,\rf{batchelor49,sreen97}).
Non-Gaussian probability distribution of
turbulent quantities are a footprint of intermittency that produces the
heavy tails of the distribution functions~\citep{frisch, mini10}.

\begin{figure}
\centering
\includegraphics[width=\textwidth]{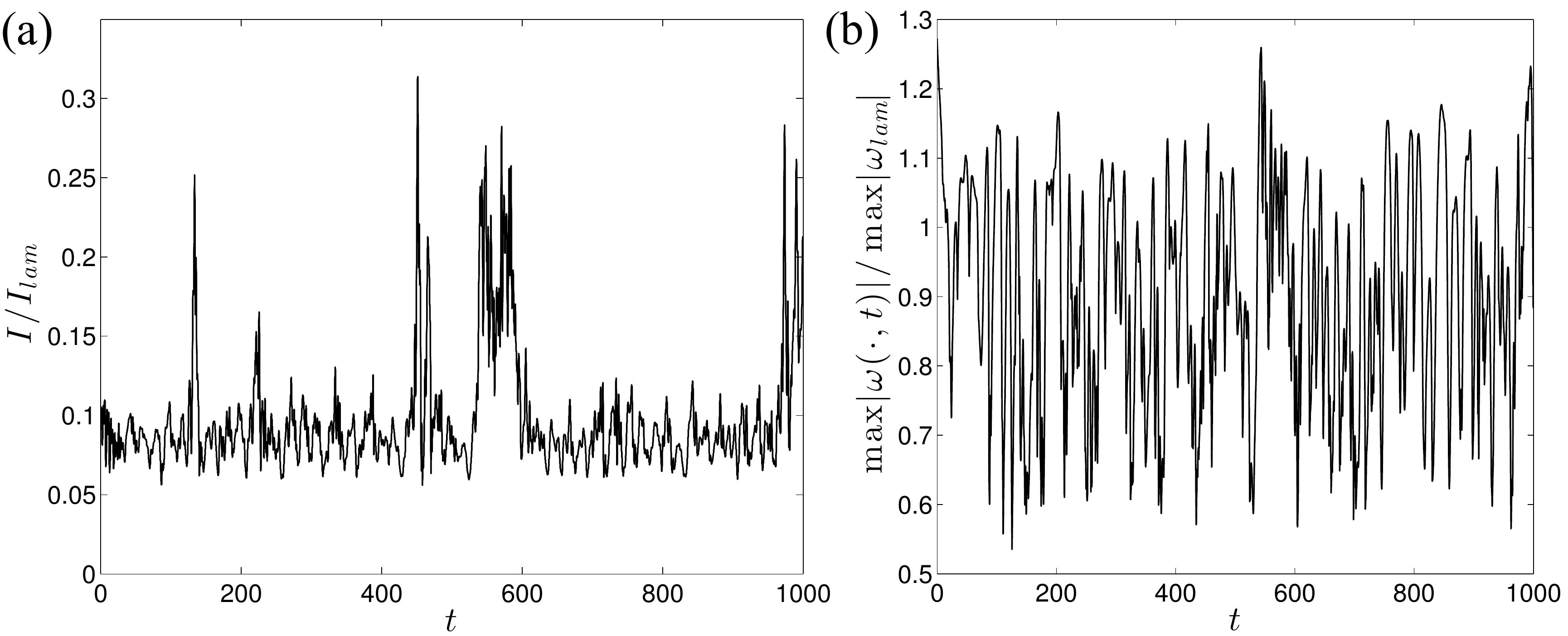}
\caption{(a) The energy input $I$ as a function of time normalized by the energy
input of the laminar state $I_{lam}=1.25$.
(b) The spatial maximum of
vorticity magnitude, i.e. $\max_{\vc x\in\mathbb T^2}|\omega(\vc x,t)|$, normalized by its value
corresponding to the laminar state, i.e.,
$\max_{\vc x\in\mathbb T^2}|\omega_{lam}(\vc x)|=10$.
}
\label{fig:It}
\end{figure}

The short lifetime of the intermittent bursts is better seen in Fig.~\ref{fig:It}
(left panel) where the normalized energy input $I/I_{lam}$ is shown
as a function of time. The time series for the energy dissipation
(not shown here) is very similar, except that the intermittent bursts of
the dissipation occur with a short delay of $0.5$ to $1.5$ time units
relative to the energy input bursts. This suggests that, once in a while,
the turbulent velocity field mostly aligns with the external forcing $\sin(nx_2)\vc e_1$
resulting in the growth of the energy input which kicks the
trajectory out of its ergodic sea. After a short time delay, the energy
dissipation also increases, bringing the trajectory back to the ergodic regime.

Spatial intermittency is
also a characteristic of turbulent fluid flow which refers to
unusually large velocity (or vorticity) amplitudes occurring in a relatively small subset of the
physical domain (see, e.g.,\rf{Kuo71,schneider04,farge09}).
Although temporal and spatial
intermittencies are sometimes conflated in the literature,
the relation between the two
is not well-understood~\citep{gibbon03}. For our turbulent trajectory, in fact, an appreciable
correlation
between them was not found. Figure~\ref{fig:It}(b), for instance, shows
that the normalized maximum vorticity amplitude oscillates rapidly throughout the simulation
time, exhibiting no clear correlation with the energy input.

Focusing on the temporal intermittency, we first review the dynamical systems
perspective of this phenomena.


\subsection{A dynamical systems perspective on temporal intermittency}\label{sec:inter_DS}
Although the Navier--Stokes equations generate an infinite-dimensional dynamical system,
it is believed that due to the dissipative term $\nu\Delta\vc u$, its
solutions converge exponentially fast to a
finite-dimensional, invariant set, usually referred to as the \emph{inertial manifold}
(see~\cite{const12} for the rigorous definition). The existence of the
inertial manifold for the Navier--Stokes equation, in its most general form,
is an open mathematical problem. In
practice, however, its existence is often assumed. In fact, this assumption
underlies the finite Galerkin truncations used in computations~\citep{foias88}.

\begin{figure}
\centering
\includegraphics[width=.85\textwidth]{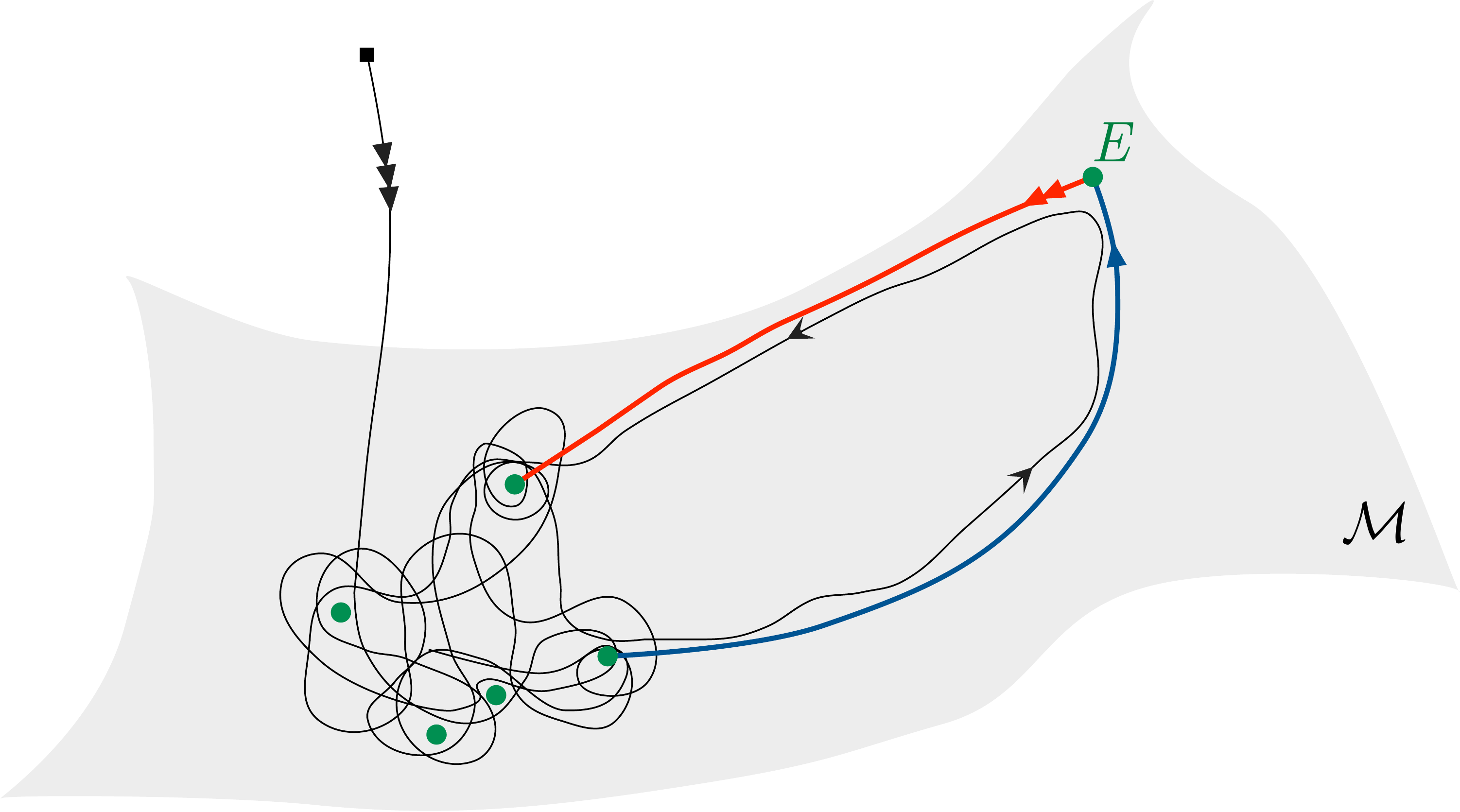}
\caption{An initial condition (black square) decays rapidly to the
inertial manifold $\mathcal M$ where the dynamics is governed by the (relative) equilibria,
(relative) periodic orbits, and portions
of their stable/unstable manifolds that
lie within the inertial manifold. The green dots represent equilibria.
Highly unstable invariant solutions (e.g. the equilbrium E)
are rarely visited by a generic trajectory.}
\label{fig:IM}
\end{figure}

Some (relative) equilibria
and (relative) periodic orbits and portions of their stable and unstable manifolds
belong to the inertial manifold.
For large enough Reynolds numbers, these invariant solutions
are typically unstable. A generic turbulent trajectory visits the neighborhood of an invariant
solution
for a finite time before it is repelled along its unstable manifold towards
the neighborhood of another invariant solution~\citep{ruelle91,HGCV09}.
This process continues indefinitely
in a somewhat unpredictable fashion, thereby causing the complex temporal behavior of
turbulent trajectories (see Fig.~\ref{fig:IM} for an illustration).

From this perspective, intermittent episodes are viewed as close passages of the
turbulent trajectory to invariant solutions that reside in a `less accessible' region of the
inertial manifold or, more precisely, the attractor~\citep{holmes93}.
As depicted in Fig.~\ref{fig:IM}, such passages are viable along the heteroclinic connections
between the invariant solutions~\citep{holmes92,Holmes96}.

To characterize the less accessible regions of the attractor, one naturally needs to
answer the following question:
How frequently is an invariant solution visited by generic turbulent trajectories?
There is no straightforward, a priori answer to this
question~\cite[Chapter 23]{DasBuch}. There are, however,
some characteristics of the invariant solutions that are relevant. In Tables~\ref{tab:EQ}
and~\ref{tab:TW}, for instance, we report the dimension of the linear unstable manifold
(i.e., $\dim E^u$) of each invariant solution.
The invariant solutions $E_1$, $E_4$, $E_7$, $T_1$, $T_2$, $T_3$, $T_7$,
$T_8$ and $T_9$ that reside close to the ergodic sea have at most $10$
linearly unstable eigenmodes. The remaining invariant solutions have at
least $13$ unstable eigenmodes and seem to reside further away from the
ergodic sea.

In Tables~\ref{tab:EQ}
and~\ref{tab:TW}, we also report the stability exponent $\mu_1+i\omega_1$ of the most unstable
eigenmode of each invariant solution. It is tempting to assert that the
invariant solutions with larger $\mu_1$ are less likely to be visited by a generic turbulent
trajectory. This is, however, not the case. For instance, we have $\mu_1=0.62697$ for equilibrium
$E_4$ and $\mu_1=0.61189$ for equilibrium $E_8$. Equilibrium $E_4$ is located in the heart of
the ergodic sea and is, in fact, visited by the turbulent trajectory quite often. Equilibrium $E_8$,
in spite of having a similar stability exponent, is rarely visited by the turbulent trajectory.
The crucial difference between these two equilibria is the dimension of the linear unstable
manifold which are $\dim E^u=5$ for $E_4$ and $\dim E^u=17$ for $E_8$.

In retrospect, the lack of correlation between the stability exponent and the frequency
of visitations by the turbulent trajectory is to be expected. The stability exponent
of an invariant solution is a
local quantity. As such, for it to be meaningful, the turbulent trajectory should
already be in the vicinity of the invariant solution. Once there,
the stability exponent $\mu_1$ determines how quickly the trajectory will leave
the neighborhood.

At any rate, neither the dimension of the unstable manifold nor the stability exponents
of an invariant solution decisively determine the frequency at which its
neighborhood is visited by a turbulent trajectory. Therefore,
we take a more direct approach to quantify intermittency. Namely, we measure
the $L^2$ distance between the turbulent trajectory and the computed invariant solutions.

The Kolmogorov equation is equivariant under a one-parameter
family of continuous symmetries and a number of discrete symmetries~\citep{sirovich87}.
This implies that
each solution $\vc u(t)$ has infinitely many
equivalent copies. Therefore, when measuring the `distance' between two states
one needs to make an informed choice among the equivalent copies of each state.
This necessitates a discussion on the symmetries of the Kolmogorov flow.

\subsection{Symmetries of Kolmogorov flow}\label{sec:sym}
The Kolmogorov equation~\eqref{eq:kolm} is equivariant with respect to $4n$ discrete symmetries and
a continuous
translational symmetry~\citep{sirovich87}. We denote the complete set of such symmetries by
\begin{subequations}
\begin{alignat}{3}
\left(\mathcal T_{\ell}\vc u\right)(x_1,x_2)& =\vc u(x_1+\ell,x_2)\\
\left(\mathcal R\vc u\right)(x_1,x_2)       &=-\vc u(-x_1,-x_2)  \\
\left(\mathcal S^m\vc u\right)(x_1,x_2) &=
\begin{pmatrix}
(-1)^m u_1\left((-1)^mx_1,x_2+m\pi/n\right)\\
      \quad\qquad u_2\left((-1)^mx_1,x_2+m\pi/n\right)
\end{pmatrix},
\end{alignat}
\label{eq:sym_u}%
\end{subequations}
where $m\in\{0,1,\cdots, 2n-1\}$. Here, $\mathcal T_\ell$ denotes $\ell$-shift in the
$x_1$-direction,
$\mathcal R$ denotes rotation through $\pi$ and $\mathcal S$ denotes a simultaneous $(\pi/n)$-shift
in the $x_2$-direction and a reflection in the $x_1$-direction.

One can readily verify that the above symmetry operations act on the vorticity field according to
the following rules:
\begin{subequations}
\begin{alignat}{3}
\left(\mathcal T_{\ell}\omega\right)(x_1,x_2) & =\omega(x_1+\ell,x_2) \\
\left(\mathcal R\,\omega\right)(x_1,x_2)        & =\omega(-x_1,-x_2) \\
\left(\mathcal S^m\omega\right)(x_1,x_2)      &=
(-1)^m \omega\left((-1)^mx_1,x_2+m\pi/n\right).
\end{alignat}
\end{subequations}

The glide reflection (or shift-reflect operation) $\mathcal S$ generates a cyclic group of order $2n$,
$$C_{2n}=\{e,\mathcal S,\mathcal S^2,\cdots,\mathcal S^{2n-1}\},$$
where $e$ denotes the
identity $e=\mathcal S^0$.
The rotation-through-$\pi$ operation $\mathcal R$ generates
a cyclic group of order two, $R_2=\{e,\mathcal R\}$. The complete set of discrete symmetries
of the Kolmogorov equation, therefore, is the dihedral group of order $4n$, i.e.,
$$D_{4n}=R_2\ltimes C_{2n}=\{e,\mathcal S,\cdots, \mathcal S^{2n-1},
\mathcal R,\mathcal R\mathcal S,\cdots,\mathcal R\mathcal S^{2n-1} \}.$$
Note that the operations $\mathcal R$ and $\mathcal S$ do not commute,
$\mathcal R\mathcal S\neq \mathcal S\mathcal R$. Instead, we have
$\mathcal S\mathcal R\mathcal S=\mathcal R$.

Therefore, the solutions of the Kolmogorov equation have up to $4n$ distinct but
equivalent copies due to its equivariance under the discrete symmetries $D_{4n}$.
They also have infinitely many equivalent copies due to the continuous symmetry
$\mathcal T_\ell$ for any $\ell\in[0,2\pi]$.

An invariant solution may itself have some, all or none of the symmetries of the equations.
The laminar solution $E_0$ for instance has the complete set of symmetries, i.e., $gE_0=E_0$ for
all $g\in D_{4n}$ and $\mathcal T_\ell E_0=E_0$ for all $\ell\in[0,2\pi]$. The
laminar state, therefore, has only one copy.
The traveling wave $T_7$ (see Fig.~\ref{fig:TW}), on the other hand, has no symmetries and therefore
possesses infinitely many equivalent copies. Incidentally, traveling wave $T_7$ happens to have the
lowest dimensional unstable manifold, $\dim E^u=2$, among the solutions found here.

These symmetry related copies greatly complicate the analysis of the state space of the Kolmogorov
flow. When comparing the $L^2$ distance between two states $\vc u^1$ and $\vc u^2$,
one needs to take the minimum $L^2$ distance between $\vc u^1$ and $\vc u^2$ and all their symmetry
related copies. For example, let $\vc u^2(t)$ to be a symmetry copy of
a solution $\vc u^1(t)$ such that
$\vc u^2(t)=(\mathcal T_\ell\vc u^1)(t)$ for some $\ell\in(0,2\pi)$. These two
states are equivalent and
both solve the Kolmogorov equation. However, the norm $\|\mathcal T_\ell\vc u^1-\vc u^1\|_{L^2}$ is
generally non-zero. Therefore, an appropriate norm on the state space of the
Kolmogorov flow is
\beq
\min
\|\vc u^1-g\,\mathcal T_{\ell}\vc u^2\|_{L^2}
\,,
\eeq
where the minimum is taken over all $\ell\in[0,2\pi]$ and $g\in D_{4n}$.
Evaluating this norm can be somewhat cumbersome.

Another approach is to map each state into a symmetry-invariant polynomial
basis.
For low dimensional dynamical systems with simple discrete
symmetries, such coordinates are available
analytically~\citep{GL-Gil07b}. As the dimension of the system (and/or the order of the
group) increases, the
determination of the invariant coordinates becomes quickly prohibitive~\citep{SiCvi10}.
As a result, and to the best of our knowledge,
symmetry-invariant polynomial
coordinates for the Kolmogorov flow are not known.

Here, we take an alternative approach which also proves to be insightful in analysis
of the temporal intermittency. We define the projection operator
\beq
\mathcal P\mathbf u = \frac{1}{4n}\sum_{m=0}^{2n-1}\mathcal S^m\left(\mathbf u +\mathcal R\mathbf
u\right),
\label{DnA1proj}
\eeq
which is the average over all copies of $\vc u$ given by the discrete symmetries
$D_{4n}$~\citep{DasBuch}.
The projection $\mathcal P\vc u$ is invariant under all discrete symmetries $g\in D_{4n}$
and therefore we refer to it as the \emph{symmetric part} of the state $\vc u$. The symmetric
part of vorticity $\omega$ is defined analogously.

All $4n$ symmetry copies of the state $\vc u$ have a unique projection $\mathcal P\vc
u$. Working with the symmetric part of the states, therefore, eliminates the complications
arising from the discrete symmetries.

Considering this symmetric part also has a physical
motivation. The energy $E$, dissipation $D$ and energy input $I$ defined in~\eqref{eq:IDE}
are invariant under symmetry operations. Denoting the energy input of a state $\vc u$
by $I[\vc u]$, we have $I[\vc u]=I[g\vc u]$ for $g$ being any symmetry of the Kolmogorov flow.
The same holds for energy $E$ and dissipation $D$. The particular linear form of the energy input $I[\vc u]$
implies that it is also invariant under the projection operation
$\mathcal P$, i.e. $I[\vc u]=I[\mathcal P\vc u]$.

Furthermore, we have
\beq
I[\vc u-\mathcal P\vc u]=0,
\label{eq:I_lin}
\eeq
that is, the remainder $\vc u-\mathcal P\vc u$ does not contribute to the energy input.
As discussed in Section~\ref{sec:inter_DS}, the intermittent episodes of the flow are triggered
by high energy inputs. Therefore, to study the temporal intermittency of the Kolmogorov flow, it is
sufficient to consider the symmetric part $\mathcal P\vc u$.
Note, however, that due to the quadratic form of $E$ and $D$,
the energy and dissipation of the remainder $\vc u-\mathcal P\vc u$
are generally nonzero.
\begin{figure}
\centering
\includegraphics[width=\textwidth]{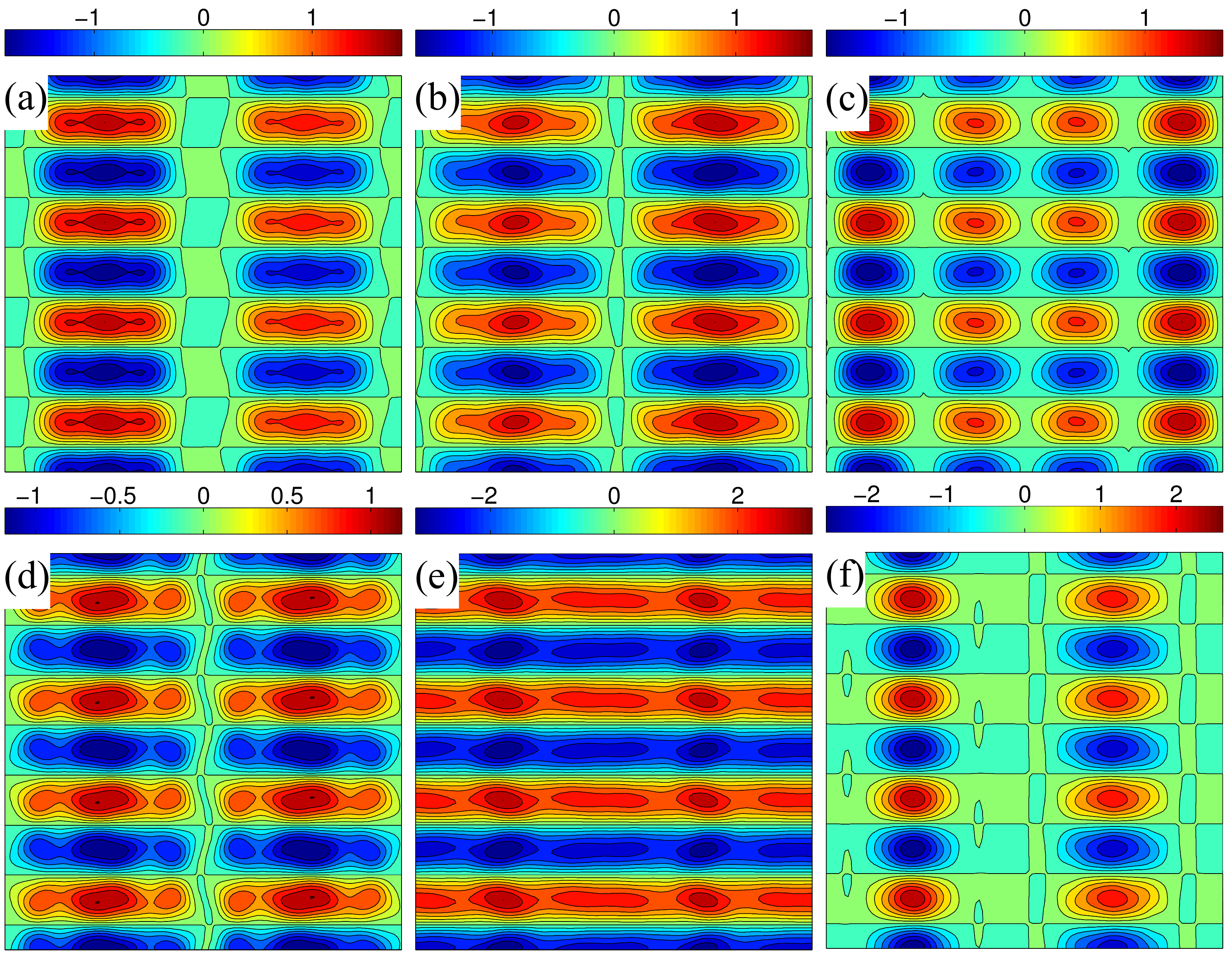}
\caption{The symmetric part of the vorticity field $\mathcal P\omega$
of the turbulent trajectory at
$t=84$ (a),
$t=94$ (b),
$t=104$ (c),
$t=124.2$ (d),
$t=134.2$ (e) and
$t=144.2$ (f).
The energy input/dissipation pairs $(I,D)$ for the states are
$(0.109,0.107)$,
$(0.115,0.114)$,
$(0.104,0.104)$,
$(0.095,0.094)$,
$(0.314,0.269)$ and
$(0.099,0.093)$, respectively.}
\label{fig:Pw_erg}
\end{figure}

The remaining continuous symmetry $\mathcal T_\ell$, is handled here by the method of
\emph{slices}~\citep{cartan35,field80,rowley03}.
This method replaces all continuous symmetry copies of a
state with a
copy that belongs to a given hypersurface called the slice. This hypersurface is such that each
group orbit $\mathcal T_\ell\vc u$
in a neighborhood of a given state
intersects the slice transversally at a unique point.
Each group orbit $\mathcal T_\ell\vc u$ is then replaced by its unique intersection with the slice.
The method of slices has only recently been used in the context of fluid
dynamics~\citep{ACHKW11,WSC15}.
We use the first-Fourier-mode implementation of this
method developed by\rf{bud15} (cf. Appendix~\ref{app:sym_fs} for more detail).

\subsection{Temporal intermittency}
Figure~\ref{fig:Pw_erg} shows the symmetric part of the vorticity field $\mathcal P\omega$
for 6 select times along the turbulent trajectory. Since the forcing wave number is
$n=4$, there are $4n=16$ discrete symmetries. As a result, the symmetric part of each state
exhibits recurring patterns that are related through discrete symmetry operations $\mathcal R$
and $\mathcal S$. In other words, knowing the symmetric part $\mathcal P\omega$ on one-sixteenth of
the domain $[0,2\pi]\times[0,2\pi]$, one can reproduce $\mathcal P\omega$ on the entire domain.
\begin{figure}
\centering
\includegraphics[width=\textwidth]{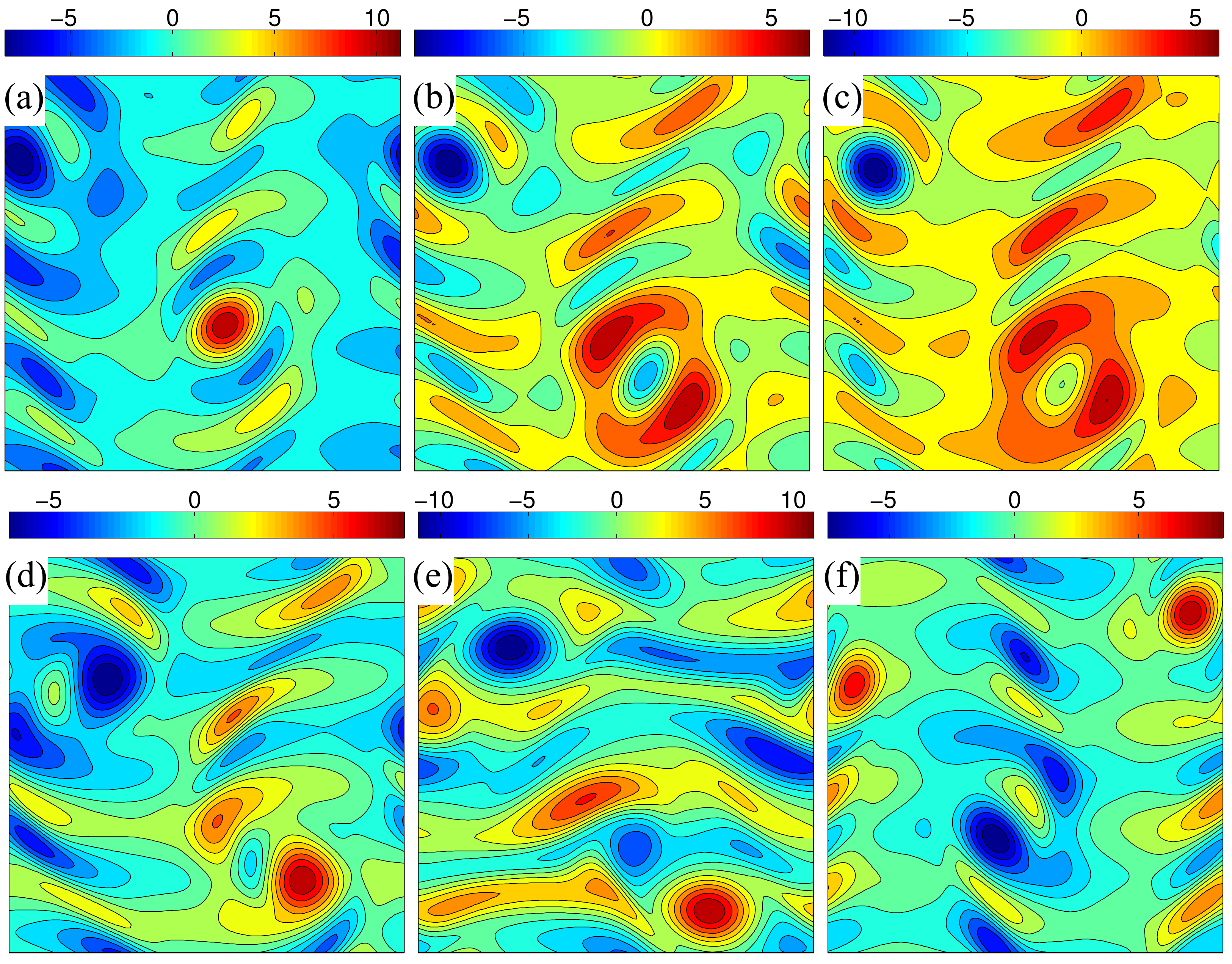}
\caption{The vorticity field $\omega$ for the snapshots of the turbulent trajectory 
shown in Fig.~\ref{fig:Pw_erg}. The panels correspond to times
$t=84$ (a),
$t=94$ (b),
$t=104$ (c),
$t=124.2$ (d),
$t=134.2$ (e) and
$t=144.2$ (f).
}
\label{fig:w_erg}
\end{figure}

Panel (e) in Fig.~\ref{fig:Pw_erg}, showing the state at time $t=134.2$,
corresponds to the first intermittent peak in Fig.~\ref{fig:It}(a). A
distinct change of topology occurs in the symmetric part of
the vorticity field as the turbulent trajectory undergoes an intermittent episode.
Before and after the episode, $\mathcal P\omega$ has at least two
distinct co-rotating vortices in each positive (or negative) vorticity band. As the trajectory
gets closer to the intermittent episode, the co-rotating vortices merge, resulting
in horizontal bands of alternating positive and negative vorticity. After the episode
(see panel (f)), these bands split again into two distinct co-rotating vortices.
This sharp distinction is not immediate from the full vorticity field $\omega$
(cf. Fig.~\ref{fig:w_erg}).
The same trend (i.e., the merger of co-rotating vortices) was observed during
the intermittent episodes occurring at later times.

\begin{figure}
\centering
\includegraphics[width=\textwidth]{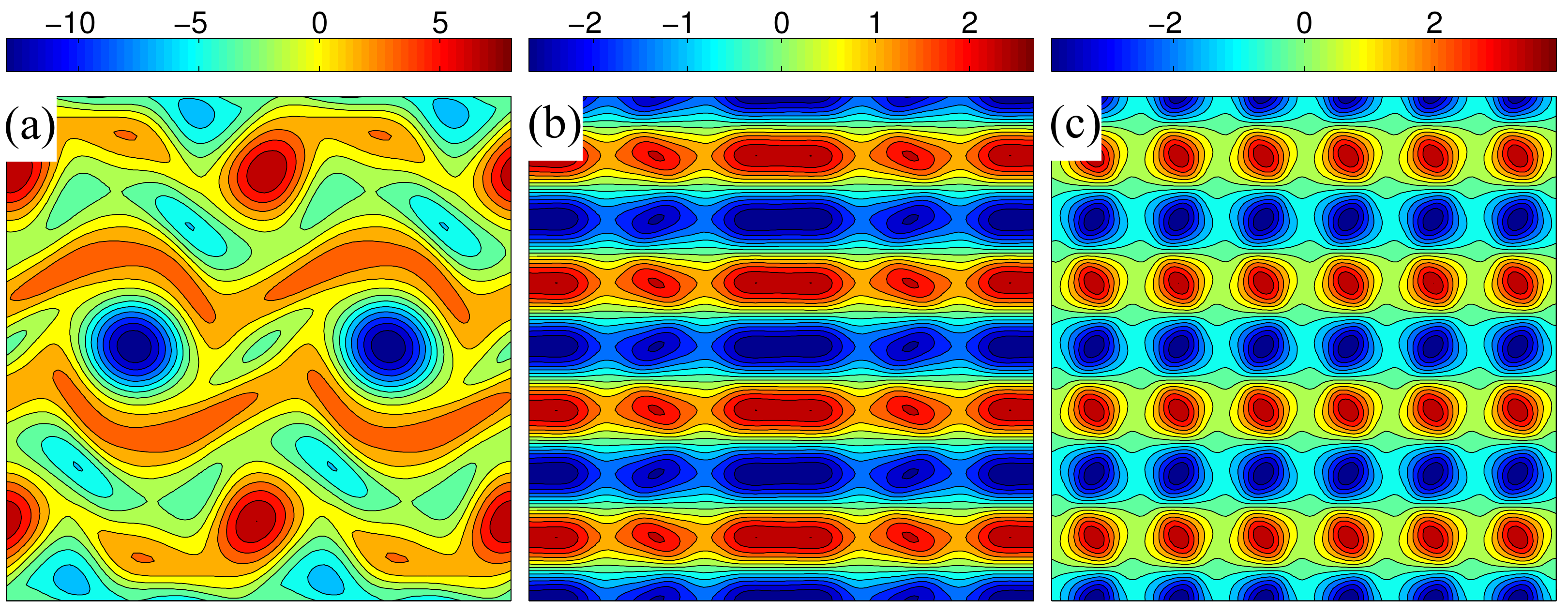}
\caption{
(a) The vorticity field $\omega$ for the equilibrium solution $E_{13}$.
(b) The symmetric part of the vorticity $\mathcal P\omega$ for the equilibrium solution $E_{13}$.
(c) The symmetric part of the vorticity $\mathcal P\omega$ for the equilibrium solution $E_{6}$.
}
\label{fig:E13}
\end{figure}

We found by inspection that the symmetric part of equilibrium solution $E_{13}$ (see Fig.~\ref{fig:E13})
is strikingly similar to that of the turbulent trajectory as it undergoes intermittency.
This is visually appreciable from comparing Fig.~\ref{fig:Pw_erg}(e) with Fig.~\ref{fig:E13}(b).
This observation suggests that close passages to equilibrium $E_{13}$ trigger the intermittent
behavior.

In Fig.~\ref{fig:E13}(c), we also show the symmetric part $\mathcal P\omega$ for
equilibrium $E_6$. While the energy input/dissipation of equilibria
$E_6$ and $E_{13}$ are very close (cf. Table~\ref{tab:EQ}), their vorticity fields are very
different. This demonstrates the fact that closeness of the energy input/dissipation of two states
does not imply their closeness in the state space.

To quantitatively examine the role of equilibrium $E_{13}$ on intermittency, we consider the $L^2$
distance between the symmetric part of each turbulent state and
the symmetric part of equilibrium $E_{13}$, i.e.,
$\|\mathcal P\omega (t)-\mathcal PE_{13}\|_{L^2}$.
To account for the continuous translational symmetry, each symmetric
part is first brought to the first-Fourier-mode slice,
as explained in Appendix~\ref{app:sym_fs}, before the distance is computed.
\begin{figure}
\centering
\includegraphics[width=.48\textwidth]{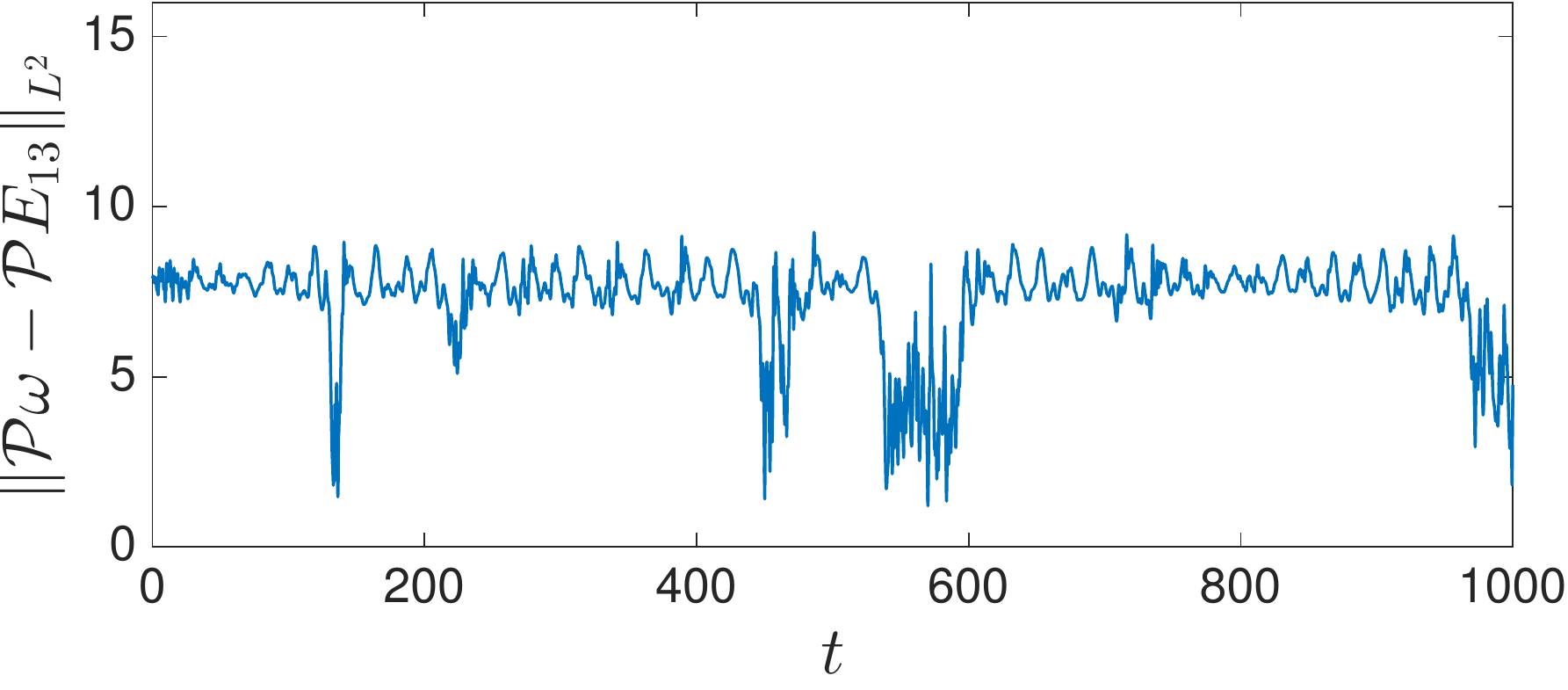}
\includegraphics[width=.48\textwidth]{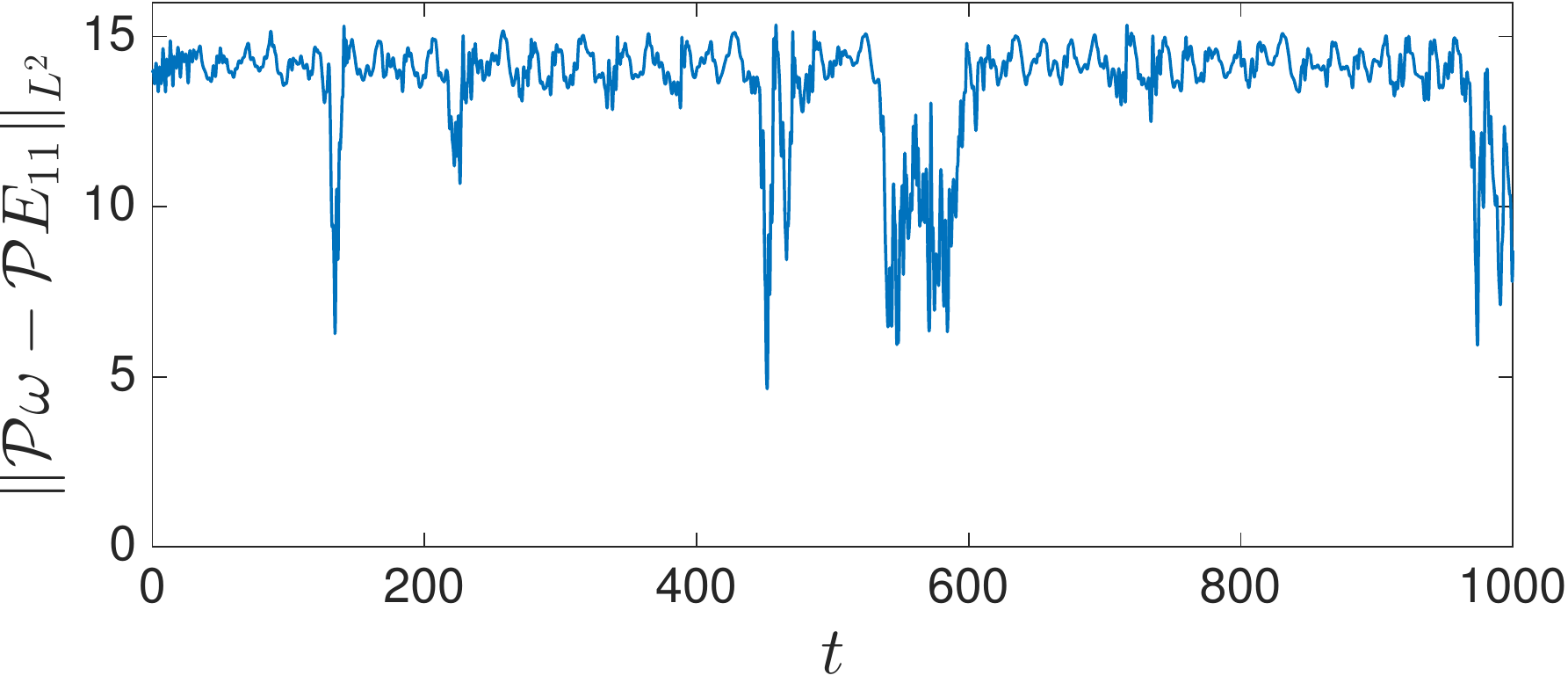}\\
\includegraphics[width=.48\textwidth]{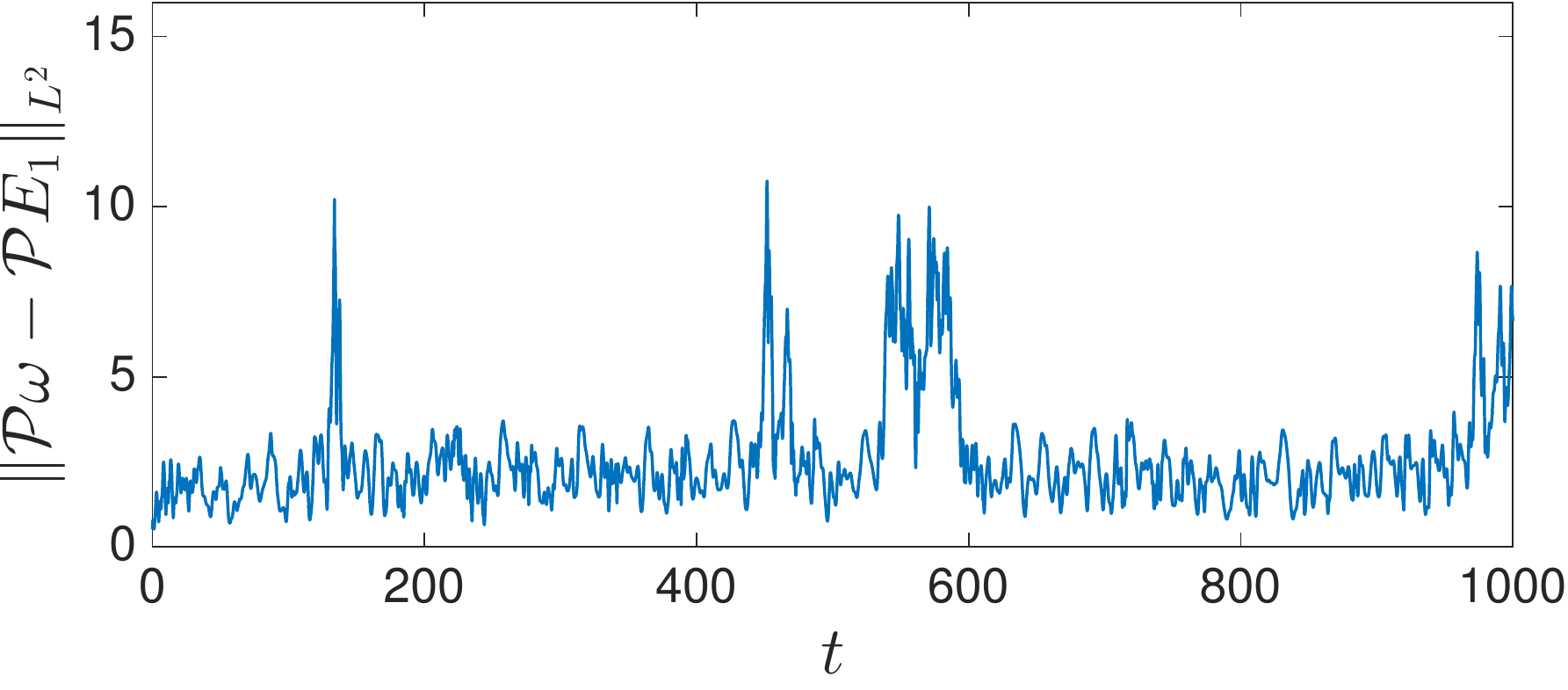}
\includegraphics[width=.48\textwidth]{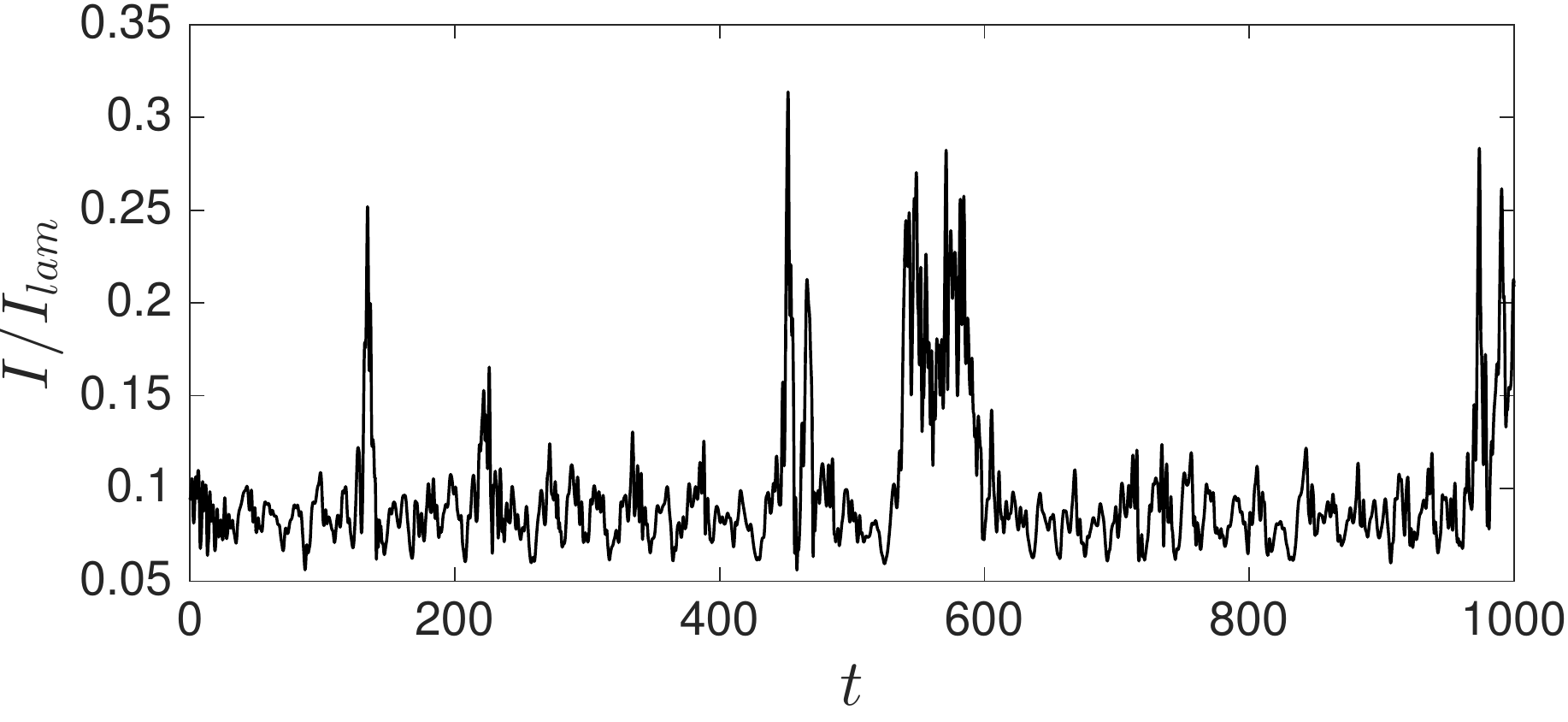}
\caption{The $L^2$ distance $\|\mathcal P\omega(t)-\mathcal PE\|_{L^2}$
between the symmetric part of the turbulent trajectory $\mathcal{P}\omega$
and the symmetric parts of the equilibria $E_{13}$ (top left), 
$E_{11}$ (top right) and $E_{1}$ (bottom left).
The normalized
energy input $I/I_{lam}$  of the turbulent
trajectory is shown for reference (bottom right; same as Fig.~\ref{fig:It}(a)).
Each distinguished peak of the energy input (i.e. intermittency) coincides with a significant
decrease in the distance from the equilibrium $E_{13}$.}
\label{fig:distE13}
\end{figure}

Figure~\ref{fig:distE13} shows the evolution of the $L^2$ distance
as a function of time. The evolution of the energy input $I$ is
also shown in the same figure. Every intermittent episode, i.e. significant peaks
in the energy input, is concurrent with a significant dip in the $L^2$ distance.
This demonstrates that the intermittent episodes correspond to
close passages of the turbulent trajectory to equilibrium $E_{13}$.

The distance between the turbulent trajectory and some of the other invariant
solutions (e.g. $E_{11}$ and $T_4$) also decreases during the intermittent episodes.
Their decrease is, however, not as dramatic as the one observed for equilibrium $E_{13}$.
For instance, the evolution of the distance from $E_{11}$ is also
shown in Fig.~\ref{fig:distE13}. The distance from $E_{11}$ fluctuates mostly around $14.0$ 
which is significantly larger that the average distance from $E_{13}$, which fluctuates around $8.0$.
During the intermittent episodes, the distance from both equilibria decreases. The distance from 
$E_{11}$ reaches $4.64$ at its minimum which is approximately $1/3$ of its average distance 
of $14.0$. The decrease in the distance from $E_{13}$ is more significant: it attains a minimum as low
as $1.21$ which is approximately $6.6$ times smaller than its average distance of $8.0$.
In contrast, the distance from equilibria residing in the ergodic sea increases during each intermittent episode.
The distance from the equilibrium $E_1$ is shown in Fig.~\ref{fig:distE13} as an example.

If the Kolmogorov flow at Reynolds number $Re=40$ possesses an ergodic attractor,
as the numerical evidence suggests (see, e.g.,\rf{PlSiFi91}),
every invariant solution embedded in the attractor will
eventually be visited by a generic turbulent trajectory.
Nonetheless, for the finite-time (and relatively short)
turbulent trajectory computed here, equilibrium $E_{13}$ appears to explain its
observed intermittencies.

Finding the invariant equilibrium and traveling wave solutions is only the first step.
A complete understanding of the Kolmogorov flow in terms of its invariant solutions
will require a detailed state space analysis of the type carried out by\rf{GHCW07} for
the plane Couette flow. Such a thorough study deserves its own treatment which will be
presented elsewhere.

\section{Conclusions and perspectives}
\label{sec:conclude}
We have  proposed and developed here a new
method for finding the equilibrium and traveling wave (relative
equilibrium) solutions of
the forced Navier--Stokes equations with periodic boundary conditions.
Namely, adjoint partial differential equations (PDEs) were derived whose (relative) equilibria
include those of Navier--Stokes equations. Furthermore, the (relative) equilibria of the adjoint
equations are asymptotically stable, and therefore, their trajectories converge to desired
invariant
solutions.

Applying this method to the Kolmogorov flow led to the discovery of several new
equilibrium and traveling wave solutions. Specifically, for Reynolds number $Re=40$,
we found a total of $24$ non-trivial equilibrium and traveling wave solutions,
where only $3$ of them were known previously~\citep{CK13}.

Some of these new solutions exhibit highly localized spatial structures
(cf. figures~\ref{fig:EQ} and~\ref{fig:TW}). Such localized structures were
previously believed to only exist in rectangular domains with
large aspect ratios~\citep{SGB10,LK14}.

One of the equilibrium solutions appears responsible for the
observed temporal intermittencies in the Kolmogorov flow. Such intermittencies manifest
themselves as sudden and short-lived bursts in the energy dissipation (and energy input)
along a generic turbulent trajectory. We showed that these bursts correspond to
close passages of the trajectory to equilibrium $E_{13}$. This supports the
dynamical systems view that such rare, extreme events are the result of heteroclinic
orbits carrying the trajectory to less accessible regions of the state space~\citep{holmes93}.

Due to the periodic boundary conditions, the adjoint PDEs can be easily integrated numerically using
a pseudo-spectral method. The spectral representation of the adjoint equations, presented
in Appendix~\ref{app:adj_fs}, exhibits a close resemblance to that of the Navier--Stokes equations.
Therefore, existing pseudo-spectral codes can be easily adapted to solve the adjoint equations.

We found the rate of convergence of the adjoint method to be rather slow. More precisely, the
trajectories of the adjoint PDEs converge rapidly to the vicinity of an invariant solution, but
further convergence takes place at a slow rate. To achieve better convergence, we proposed a
hybrid adjoint-Newton algorithm, consisting of two steps: First, the adjoint equations are
integrated from some initial condition in order to reach the neighborhood of an invariant solution.
Once in the neighborhood of the invariant solution, the standard Newton--GMRES-hook
iterations~\citep{DV04} were used to converge further to the solution. This hybrid algorithm
yielded a $100\%$ converge rate from generic initial conditions~\eqref{eq:ig},
obviating the preprocessing step for finding `good' initial guesses~\citep{DV04,CK13}.

Newton--GMRES-hook iterations are relatively expensive. Hence, a modification of
our adjoint-based method, that would eliminate the Newton-type steps altogether, is highly
desirable (see \rf{LY07}, in the context of nonlinear wave equations).

While we only considered periodic boundary conditions, the general adjoint-based approach applies
to wall-bounded turbulence such as plane Couette and pipe flows. Our preliminary analysis
(not presented here) shows that, in the presence of boundaries, the resulting adjoint equations
require more boundary conditions than the corresponding Navier--Stokes equations. This is to be
expected as the adjoint equations have higher order spatial derivatives. This calls for
a special numerical treatment of the adjoint equations for wall-bounded flows. Our results
on the Kolmogorov flow, however, show that the gain is worth the pain.

Finally, we point out that our adjoint-based method does not immediately
apply to the computation of periodic and relative periodic orbits.
They are currently found through Newton--GMRES-hook iterations
(see, e.g.,\rf{kawahara06,DV04,ACHKW11,LucKer14}) with the drawbacks discussed in the Introduction.
Alternatives include the variational method of~\rf{lanVar1} which shares the universally
convergent property of our adjoint-based method. Its computational complexity is, however,
formidable~\citep{Faz10}. More recent method of\rf{yang15} has proved promising for
unidirectional wave equations, but its feasibility for Navier--Stokes equations remains to
be explored.

\section*{Acknowledgments}
I am grateful to P.~Cvitanovi\'c for his support, helpful discussions
and a careful reading of the manuscript
that greatly improved the presentation of this work.
I am thankful to A.~P.~Willis for
his help with the implementation of Newton--GMRES-hook iterations
and his comments on an earlier version of the manuscript.
I would like to acknowledge fruitful discussions with R.~de la Llave, F.~Fedele, J.~F.~Gibson,
R.~R.~Kerswell, B.~Protas A.~Souza and L.~S.~Tuckerman.
I thank three anonymous referees for their comments and suggestions that 
improved the presentation of the results.
M.~F.\ thanks the family of G.~Robinson,~Jr.\ for J.~Ford Fellowship
support.

\section*{Supplementary material}
A tutorial, with the accompanying MATLAB code, is available  at\\
\href{https://farazmand.wordpress.com/2015/12/18/adjoint/}{https://farazmand.wordpress.com/2015/12/18/adjoint/}.

\appendix
\section{Derivation of the adjoint equations for equilibria}\label{app:proof_adj}
We start by redefining the vector $\vc F_{\vc 0}$ by adding the term $\bnabla\cdot\vc u$ as
a new element; that is
\beq
\vc F_{\vc 0}(\vc q)=
\begin{pmatrix}
-\vc u\cdot\bnabla\vc u -\bnabla p +\nu\Delta\vc u+\vc f\\
\bnabla\cdot \vc u	
\end{pmatrix},
\label{eq:ext_rhs}
\eeq
where $\vc q=(\vc u,p)$ and with the understanding that for divergence-free vector
fields, the last row of $\vc F_{\vc 0}$ is identically zero. This twist in the notation
proves useful below, where we compute the adjoint of the G\^ateaux derivative within
the space of divergence-free vector fields.
The G\^ateaux derivative of $\vc F_{\vc 0}$ is given by
\beq
\pmb{\mathcal L}_{\vc 0}(\vc q;\vc q')=
\begin{pmatrix}
-\vc u'\cdot\bnabla\vc u -\vc u\cdot\bnabla\vc u' -\bnabla p' +\nu\Delta\vc u' \\
\bnabla\cdot \vc u'
\label{eq:gateaux_NS}
\end{pmatrix}
\eeq
where $\vc q'=(\vc u',p')$ with $\bnabla\cdot\vc u'=0$.

The key part of deriving the adjoint equation~\eqref{eq:adjPDE_NS} is to
find the adjoint operator $\pmb{\mathcal L}^\dagger_{\vc 0}$. Its derivation is standard
and can be found in the literature on adjoint-based optimal control (see, e.g.,\rf{gunzburger}; our
notation is closer to\rf{faraz11}).
The difference is that, in optimal control, one seeks to minimize
a cost functional with the constraint that the Navier--Stokes equations are
satisfied. Here instead, the only constraint is the divergence-free
condition and we seek to find the states $\vc q=(\vc u,p)$ such that
$\|\vc F_{\vc 0}(\vc q)\|_{\mathcal A}=0$. We derive the adjoint
with respect to the $L^2$ inner product first. The adjoint
with respect to the more general inner product~\eqref{eq:innp_H-1} follows easily.

Let the function space $\mathcal H$ be the space of square integrable functions
$\vc q=(\vc u,p)$ such that the $\vc u$ component is divergence-free. More precisely,
\beq
\mathcal H=\{\vc q=(\vc u,p)\in L^2(\mathcal D)\, |\, \bnabla\cdot\vc u=0 \}.
\label{eq:L2divfree}
\eeq
We define the usual $L^2$ inner product on $\mathcal H$, i.e., for any $\vc q,\vc
q'\in\mathcal H$ we
define
\beq
\langle \vc q,\vc q'\rangle_{L^2}=\int_{\mathcal D} \left[\vc u(\vc x,t)\cdot \vc
u'(\vc x,t)+p(\vc x,t)p'(\vc x,t)\right]\mbox d \vc x
\label{eq:innp_l2}
\eeq

The adjoint of the linear operator~\eqref{eq:gateaux_NS} with respect to the inner product
\eqref{eq:innp_l2} on the function space $\mathcal H$~\eqref{eq:L2divfree} is given by
\beq
\pmb{\mathcal L}_{\vc 0}^\dagger(\vc q;\vc q'')=
\begin{pmatrix}
\left[\bnabla\vc u''+(\bnabla\vc u'')^\top\right]\vc u-\bnabla p''+\nu\Delta \vc u''\\
\bnabla\cdot \vc u''
\end{pmatrix},\quad\forall\, \vc q,\vc q''\in \mathcal H,
\label{eq:adj_NS}
\eeq
where $\top$ denotes matrix transposition.
As we are restricted to the space of divergence free vector fields, the last component
of the adjoint~\eqref{eq:adj_NS} is identically zero, i.e., $\bnabla\cdot\vc u''=0$.

This expression follows directly from definition~\eqref{eq:adj_def} of the adjoint and
Eq.~\eqref{eq:gateaux_NS}. Namely, substituting~\eqref{eq:gateaux_NS} into~\eqref{eq:adj_def}, we have
\begin{align}
\langle \pmb{\mathcal L}_{\vc 0}(\vc q;\vc q'),\vc q''\rangle_{L^2}&=
\int_{\mathcal D}
\begin{pmatrix}
-\vc u'\cdot\bnabla\vc u-\vc u\cdot\bnabla\vc u'
-\bnabla p'+\nu\Delta \vc u'\\
\bnabla\cdot\vc u'
\end{pmatrix}
\cdot
\begin{pmatrix}
\vc u''\\
p''
\end{pmatrix}\mathrm d\vc x,
\nonumber\\
 &= \int_{\mathcal D}\big[\big(\left(\bnabla\vc u''+\bnabla\vc u''^\top\right)\vc u -\bnabla p''
+\nu\Delta \vc u''\big)\cdot \vc u'+(\bnabla\cdot \vc u'')p'\big]\mathrm{d}\vc x,
\nonumber\\
 &=\int_{\mathcal D}
\begin{pmatrix}
\left(\bnabla\vc u''+\bnabla\vc u''^\top\right)\vc u -\bnabla p''
+\nu\Delta \vc u''\\
\bnabla\cdot \vc u''
\end{pmatrix}
\cdot
\begin{pmatrix}
\vc u'\\
p'
\end{pmatrix}
\mathrm{d}\vc x,
\nonumber\\
 &=\langle\vc q', \pmb{\mathcal L}_{\vc 0}^\dagger(\vc q;\vc q'')\rangle_{L^2},
\label{eq:adj_NS_l2}
\end{align}
where the first line is the definition of the $L^2$ inner product and the second line follows from
integration by parts. Note that the boundary terms vanish due to the periodic boundary conditions.
Only one of these integration by parts is not straightforward, which we
outline below.
\begin{align*}
\int_{\mathcal D}(-\vc u'\cdot\bnabla\vc u-\vc u\cdot\bnabla\vc u')\cdot\vc u''\,\mathrm d\vc x &=
\int_{\mathcal D}(-u''_iu'_j\partial_ju_i-u''_iu_j\partial_ju'_i)\,\mathrm d\vc x,\\
&=\int_{\mathcal D}(u_i\partial_j(u''_iu'_j)+u'_i\partial_j(u''_iu_j))\,\mathrm d\vc x,\\
&=\int_{\mathcal D}(u_iu'_j\partial_ju''_i+u'_iu_j\partial_ju''_i)\,\mathrm d\vc x,\\
&=\int_{\mathcal D}(u_ju'_i\partial_iu''_j+u'_iu_j\partial_ju''_i)\,\mathrm d\vc x,\\
&=\int_{\mathcal D}\big[\left(\bnabla\vc u''+\bnabla\vc u''^\top\right)\vc u\big]\cdot \vc
u'\,\mathrm d\vc x,
\end{align*}
where we used the summation notation on repeated indices and the fact that $\vc u$ and $\vc u'$
are divergence-free: $\partial_ju_j=\partial_ju'_j=0$.

Identity~\eqref{eq:adj_NS_l2} holds for general $\vc q,\vc q',\vc q''\in\mathcal H$.
Along the trajectories of the adjoint descent PDE, we have
$\vc q=(\vc u,p)$,
$\vc q'=(\partial_\tau\vc u,\partial_\tau p)$ and
$\vc q'':=(\vc u'',p'')=\vc F_{\vc 0}(\vc q)$ (cf.
Eq.~\eqref{eq:norm_ev2}). This yields,
\beq
\langle\vc q', \pmb{\mathcal L}_{\vc 0}^\dagger(\vc q;\vc q'')\rangle_{L^2}=
\int_{\mathcal D}\left\{\partial_\tau \vc u\cdot\left( \left[\bnabla\vc u''+(\bnabla\vc
u'')^\top\right]\vc u-\bnabla
p''+\nu\Delta \vc u'' \right)+(\bnabla\cdot\vc u'')\partial_\tau p\right\}\mathrm d\vc x.
\eeq
Therefore, equation~\eqref{eq:adjPDE_NS_1} ensures that the above inner product is
negative semi-definite. This together with identity~\eqref{eq:norm_ev2} ensures that
the norm $\|\vc F_{\vc 0}(\vc q(\tau))\|_{L^2}$ decreases along the trajectories of the adjoint PDE.

Due to the divergence-free condition, the term $(\bnabla\cdot\vc u'')\partial_\tau p$
vanishes. Therefore, no independent evolution equation is obtained for the pressure $p$.
As in the case of the Navier--Stokes equation, the pressure terms simply
enforce the divergence-free conditions~\eqref{eq:adjPDE_NS_2}.

The adjoint descent with respect to the more general inner product~\eqref{eq:innp_H-1}
is obtained similarly, the only difference being that
$\vc q''$ is replaced by $\mathcal A\vc q''$.

\section{Derivation of the adjoint equation for traveling waves}\label{app:adj_tw}
As in Appendix~\ref{app:proof_adj}, we redefine $\vc F_{\vc c}$ as
\beq
\vc F_{\vc c}(\vc q)=
\begin{pmatrix}
-\vc u\cdot\bnabla\vc u -\bnabla p +\nu\Delta\vc u+\vc f+\vc c\cdot\bnabla \vc u \\
\bnabla\cdot \vc u	
\end{pmatrix},
\eeq
where $\vc q=(\vc u,p)$ and
\beq
\vc F_{\vc c}(\vc q)=\vc F_{\vc 0}(\vc q)+
\begin{pmatrix}
\vc c\cdot \bnabla\vc u\\
0
\end{pmatrix},
\eeq
with $\vc F_{\vc 0}$ defined in~\eqref{eq:ext_rhs}. Due to the linearity of the second term,
the G\^ateaux derivative of $\vc F_{\vc c}$ is
\beq
\pmb{\mathcal L}_{\vc c}(\vc q;\vc q')=\pmb{\mathcal L}_{\vc 0}(\vc q;\vc q')+
\begin{pmatrix}
\vc c\cdot \bnabla\vc u'\\
0
\end{pmatrix},
\eeq
where $\vc q'=(\vc u',p')$ and $\pmb{\mathcal L}_{\vc 0}$ is
defined by~\eqref{eq:adj_NS}.

Analogous to anlysis of Appendix~\ref{app:proof_adj}, the adjoint $\pmb{\mathcal L}_{\vc
c}^\dagger$ can be shown to be given by
\beq
\pmb{\mathcal L}_{\vc c}^\dagger(\vc q;\vc q'')=\pmb{\mathcal L}_{\vc 0}^\dagger(\vc q;\vc q'')
-
\begin{pmatrix}
\vc c\cdot\bnabla\vc u''\\
0
\end{pmatrix},
\eeq
where $\vc q''=(\vc u'',p'')$.

We would like to define the adjoint descent equation in such a way that, along its trajectories
$(\vc q(\tau),\vc c(\tau))$, the residue $\|\vc F_{\vc c(\tau)}(\vc q(\tau))\|_{\mathcal A}$
decreases monotonically. Taking derivative with respect to the fictitious time $\tau$, we obtain
\begin{align}
\partial_\tau\|\vc F_{\vc c}(\vc q)\|^2_{\mathcal A}&=
\langle\pmb{\mathcal L}_{\vc c}(\vc q;\vc q'),\mathcal A\vc F_{\vc c}(\vc q)\rangle_{L^2}+
\left\langle
\begin{pmatrix}
\dot{\vc c}\cdot\bnabla\vc u\\
0
\end{pmatrix},
\mathcal A\vc F_{\vc c}(\vc q)
\right\rangle_{L^2}\nonumber\\
&=\langle\vc q',\pmb{\mathcal L}_{\vc c}^\dagger(\vc q;\mathcal A\vc F_{\vc c}(\vc q))\rangle_{L^2}+
\left\langle
\dot{\vc c}\cdot\bnabla\vc u,
\tilde{\vc u}''
\right\rangle_{L^2}\nonumber\\
&=\langle\vc q',\pmb{\mathcal L}_{\vc c}^\dagger(\vc q;\mathcal A\vc F_{\vc c}(\vc q))\rangle_{L^2}+
\dot{\vc c}\cdot\int_{\mathcal D}(\bnabla\vc u)^\top
\tilde{\vc u}''\mathrm d\vc x,
\end{align}
where $\vc q'=\partial_\tau \vc q$ and $\vc u''$ is given by~\eqref{eq:u''_tw}. The adjoint set
of equations~\eqref{eq:adjPDE_NS_H-1_tw} is designed in such a way that the
right hand side of the above equation is always negative, resulting in
monotonic decrease in the residue along its solutions $(\vc u(\tau),p(\tau))$.

\section{The adjoint equations in the Fourier space}\label{app:adj_fs}
In this Appendix, we derive the spectral representation of the adjoint descent
equations~\eqref{eq:adjPDE_NS} and~\eqref{eq:adjPDE_NS_H-1_tw}. As shorthand notation,
we define
\beq
\vc N:=\left[\bnabla\tilde{\vc u}''+(\bnabla\tilde{\vc u}'')^\top\right]\vc u,
\eeq
and denote Fourier transforms with a \emph{hat} sign. Then one can write
equation~\eqref{eq:adjPDE_NS_1}, in the Fourier space as
\beq
\partial_\tau\widehat{\vc u}(\vc k)=-\left(\vc I-\frac{\vc k\otimes \vc k}{|\vc k|^2}\right)
\widehat{\vc N}(\vc k)+\nu |\vc k|^2\,\widehat{\tilde{\vc u}''}(\vc k),
\label{eq:adjPDE_fs}
\eeq
where $\vc I$ is the identity matrix and $\vc k\otimes\vc k$ denotes the tensor product.
Note that the pressure $p''$ is eliminated using
identity~\eqref{eq:p''} which implies
$$\widehat{p''}(\vc k)=\frac{-i\,\vc k\cdot \widehat{\vc N}(\vc k)}{|\vc k|^2}.$$

For $\widehat{\tilde{\vc u}''}$, using definition~\eqref{eq:Afs}, we have
$$\widehat{\tilde{\vc u}''}(\vc k)=\frac{\widehat{\vc u}''}{1+|\vc k|^2}.$$
The term $\widehat{\vc u}''$ is in turn computed from identity~\eqref{eq:u''} as
\beq
\widehat{\vc u}''(\vc k)=-\left(\vc I-\frac{\vc k\otimes \vc k}{|\vc k|^2}\right)
\widehat{\vc R}(\vc k)-\nu |\vc k|^2\,\widehat{\vc u}(\vc k)+\widehat{\vc f}(\vc k),
\label{eq:NS_fs}
\eeq
where $\vc R$ is the shorthand notation for the nonlinear term
$$\vc R:=\vc u\cdot\bnabla\vc u.$$

The nonlinear terms $\vc R$ and $\vc N$ are computed by the pseudo-spectral method, i.e.,
the differentiations are carried out in the Fourier space and the products are computed in the
physical space.

Note that the spectral representation~\eqref{eq:adjPDE_fs} of the adjoint
equation closely resembles that of the Navier--Stokes equation~\eqref{eq:NS_fs}.
Therefore, existing pseudo-spectral codes for Navier--Stokes can easily
be adapted to solve the adjoint PDEs

The adjoint descent for the traveling waves is computed similarly,
except that the term
$i\,(\vc c\cdot\vc k)\widehat{\tilde{\vc u}''}(\vc k)$
is added to
the right-hand-side of equation~\eqref{eq:adjPDE_fs} accounting for the term
$\vc c\cdot\bnabla\tilde{\vc u}''$ in~\eqref{eq:adjPDE_NS_H-1_tw}. More precisely, the
adjoint descent equation for the traveling waves in the Fourier space reads
\beq
\partial_\tau\widehat{\vc u}(\vc k)=-\left(\vc I-\frac{\vc k\otimes \vc k}{|\vc k|^2}\right)
\widehat{\vc N}(\vc k)+\nu |\vc k|^2\,\widehat{\tilde{\vc u}''}(\vc k)+
i\,(\vc c\cdot\vc
k)\widehat{\tilde{\vc u}''}(\vc k),
\label{eq:adjPDE_tw_fs}
\eeq
with $\tilde{\vc u}''=\mathcal A\vc u''$ and
\beq
\widehat{\vc u}''(\vc k)=-\left(\vc I-\frac{\vc k\otimes \vc k}{|\vc k|^2}\right)
\widehat{\vc R}(\vc k)-\nu |\vc k|^2\,\widehat{\vc u}(\vc k)+\widehat{\vc f}(\vc k)+
i\,(\vc c\cdot\vc
k)\widehat{\vc u}(\vc k).
\eeq

\section{Symmetry actions in the Fourier space}\label{app:sym_fs}
The symmetry operations~\eqref{eq:sym_u} can be readily implemented
in the Fourier space. The following transformations follow directly from
the definition of the Fourier transform. Denoting the Fourier modes of the velocity
field $\vc u$ by $\widehat{\vc u}(\vc k)$ with $\vc k=(k_1,k_2)\in\mathbb Z^2$, we have
\beq
\vc u(x_1,x_2)=\sum_{k_1\in\mathbb Z}\sum_{k_2\in\mathbb Z}
\widehat{\vc u}(k_1,k_2)e^{ik_1x_1}e^{ik_2x_2}.
\label{eq:dft}
\eeq

Therefore, the shift operation $\mathcal T_\ell$ satisfies
\beq
\left(\mathcal T_\ell\vc u\right)(x_1,x_2)=\vc u(x_1+\ell,x_2)=
\sum_{k_1\in\mathbb Z}\sum_{k_2\in\mathbb Z}\widehat{\vc
u}(k_1,k_2)e^{ik_1\ell}e^{ik_1x_1}e^{ik_2x_2},
\eeq
implying that the action of the continuous symmetry $\mathcal T_\ell$ on the Fourier mode
$\widehat{\vc u}(k_1,k_2)$ is multiplicative, such that
\beq
\widehat{\mathcal T_\ell \vc u}(k_1,k_2)=
e^{ik_1\ell}\widehat{\vc u}(k_1,k_2).
\label{eq:shift_fs}
\eeq

Similarly, for the rotation through $\pi$ symmetry $\mathcal R$, we have
\beq
\widehat{\mathcal R \vc u}(k_1,k_2)=-\widehat{\vc u}(-k_1,-k_2)=
-\left[\widehat{\vc u}(k_1,k_2)\right]^\ast,
\eeq
where the superscript $\ast$ denotes complex conjugation and the last identity follows from
the fact that $\vc u$ is real valued.

Finally, the shift-reflect symmetry $\mathcal S$ acts on the Fourier modes through
\beq
\widehat{\mathcal S^m\vc u}(k_1,k_2) =
e^{ik_2\frac{m\pi}{n}}
\begin{pmatrix}
(-1)^m \widehat u_1\left((-1)^mk_1,k_2\right)\\
      \quad\qquad \widehat u_2\left((-1)^mk_1,k_2\right)
\end{pmatrix}.
\eeq

To reduce the continuous symmetry,
we use the first-Fourier-mode slice of\rf{bud15}.
It follows from~\eqref{eq:shift_fs} that the shift $\mathcal T_\ell$ acts on the Fourier mode
$(k_1,k_2)=(1,0)$ of the vorticity through
\beq
\widehat{\omega}(1,0)\mapsto\widehat{\omega}(1,0)e^{i\ell}=
|\widehat{\omega}(1,0)|e^{i(\phi(1,0)+\ell)},
\eeq
where $\phi(k_1,k_2)\in(-\pi,\pi]$ denotes the \emph{principal value} of
the phase of mode $(k_1,k_2)$.
To bring the vorticity field to the first-Fourier-mode slice, the shift value $\ell$ is chosen
such that $\phi(1,0)+\ell=0$. More precisely, a given vorticity field $\omega$ is replaced with its
symmetry-equivalent copy through the transformation
\beq
\widehat{\omega}(k_1,k_2)\mapsto e^{-ik_1\phi(1,0)}\widehat{\omega}(k_1,k_2).
\eeq
As a result, the mode $\widehat{\omega}(1,0)$ of the symmetry-reduced vorticity field
has vanishing imaginary part.


\end{document}

%% file: setupAdjoint.tex
            \newif\ifdraft \newif\ifcolorfigs \newif\ifJFM
            \drafttrue \colorfigstrue \JFMtrue

\ifJFM
  \usepackage{upmath}
\fi

\usepackage{ifpdf}
\ifpdf
  \usepackage[pdftex]{color,graphicx}
  \usepackage{epstopdf}
\else
  \usepackage[dvips]{color,graphicx}
\fi

\usepackage{float}
\usepackage{natbib}
\usepackage{amsmath,amsfonts,amssymb,amsbsy,amscd,amsgen}
\usepackage{url}
\usepackage{subfigure}
\usepackage{ifthen}
\usepackage[table]{xcolor}
\graphicspath{{../figs/}{../Fig/}}  
\usepackage[ruled,vlined,norelsize]{algorithm2e}

\usepackage[active]{srcltx}

\ifcolorfigs
  \usepackage[pdftex,colorlinks,citecolor=black]{hyperref} 
\else
\fi

%% file: defAdjoint.tex
\ifdraft    
   
   \newcommand{\PC}[1]{$\footnotemark\footnotetext{PC: #1}$}
   
   \newcommand{\MA}[1]{$\footnotemark\footnotetext{MA: #1}$}
   
   \newcommand{\APW}[1]{$\footnotemark\footnotetext{APW: #1}$}

   \newcommand{\MMF}[1]{$\footnotemark\footnotetext{Mohammad: #1}$}

   \newcommand{\file}[1]{$\footnotemark\footnotetext{{\bf file} #1}$}
   \newcommand{\mycomment}[2]{\noindent \textbf{\underline{#1}}: \emph{#2}}
\else   

   \newcommand{\PC}[1]{}
   \newcommand{\JFG}[1]{}
   \newcommand{\MA}[1]{}
   \newcommand{\APW}[1]{}

  \newcommand{\MMF}[1]{}
  
   \newcommand{\file}[1]{}
   \newcommand{\mycomment}[2]{}

\fi 

\ifcolorfigs            

\else

\fi

\newcommand{\rf}     [1] {~\cite{#1}}

\newcommand{\beq}{\begin{equation}}

\newcommand{\eeq}{\end{equation}}
\newcommand{\ee}[1] {\label{#1} \end{equation}}

\newcommand{\bea}{\begin{eqnarray}}
\newcommand{\eea}{\end{eqnarray}}
\newcommand{\barr}{\begin{array}}
\newcommand{\earr}{\end{array}}

\ifJFM                          

\else
\fi

\newcommand{\id}{{\ \hbox{{\rm 1}\kern-.6em\hbox{\rm 1}}}}

\newcommand{\eigExp}[1][]{
\ifthenelse{\equal{#1}{}}{\ensuremath{\lambda}}{\ensuremath{\lambda^{(#1)}}}
                        }
\newcommand{\eigRe}[1][]{
\ifthenelse{\equal{#1}{}}{\ensuremath{\mu}}{\ensuremath{\mu^{(#1)}}}
                        }
\newcommand{\eigIm}[1][]{
  \ifthenelse{\equal{#1}{}}{\ensuremath{\omega}}{\ensuremath{\omega^{(#1)}}}
            }

\newcommand{\bnabla}{\mbox{\boldmath $\nabla$}}

\newcommand{\vc}{\mathbf}